\documentclass[twocolumn,english,superscriptaddress,amsmath,amssymb,floats,prb]{revtex4-2}

\usepackage{graphicx}
\usepackage{dcolumn}
\usepackage{amssymb}
\usepackage{comment}

\usepackage{xcolor,colortbl}

\begin{document}

\title{Strain effect on the ground-state structure of Sr$_2$SnO$_4$ Ruddlesden-Popper oxides}

\author{Hwanhui Yun}
 \affiliation{Department of Chemical Engineering and Materials Science, University of Minnesota}
 
\author{Dominique Gautreau}
 \affiliation{Department of Chemical Engineering and Materials Science, University of Minnesota}
 \affiliation{School of Physics and Astronomy, University of Minnesota}

\author{K. Andre Mkhoyan}
 \affiliation{Department of Chemical Engineering and Materials Science, University of Minnesota}

\author{Turan Birol}
\email[]{tbirol@umn.edu}
 \affiliation{Department of Chemical Engineering and Materials Science, University of Minnesota}

\begin{abstract}
	Ruddlesden-Popper (RP) oxides (A$_{n+1}$B$_n$O$_{3n+1}$) comprised of perovskite (ABO$_3$)$_n$ slabs can host a wider variety of structural distortions than their perovskite counterparts. This makes accurate structural determination of RP oxides more challenging. In this study, we investigate the structural phase diagram of $n=1$ RP Sr$_2$SnO$_4$, one of alkaline earth stannates that are promising for opto-electronic applications by using group theory-based symmetry analysis and first-principles calculations. We explore the symmetry breaking effects of different dynamical instabilities, predict the energies of phases they lead to, and take into account different (biaxial strain and hydrostatic pressure) boundary conditions. We also address the effect of structural changes on the electronic structure and find that compressive biaxial strain drives Sr$_2$SnO$_4$ into a regime with wider bandgap and lower electron effective mass. 
\end{abstract}

\maketitle

\section{Introduction}

The ABO$_3$ perovskites constitute one of the most studied structural family of oxides, thanks to the plethora of interesting phenomena they host. The perovskite crystal structure is very flexible and can host a very large number of combinations of different metal cations on the A and B sites. This flexibility is in large part due to subtle crystal structural distortions (atomic displacements,) which don't break the connectivity of the corner sharing network of oxygen octahedra, but modify the coordination of the cations to compensate for under- or over-bonding. In fact, the majority of perovskite oxides have so-called oxygen octahedral rotations that reduce the cubic symmetry of the crystal structure. The cubic, high-symmetry perovskite structure is an exception observed in a small number of compounds, rather than the prototypical examples \cite{Lufaso2001}. 

\begin{figure}
    \centering
    \includegraphics[width=\columnwidth]{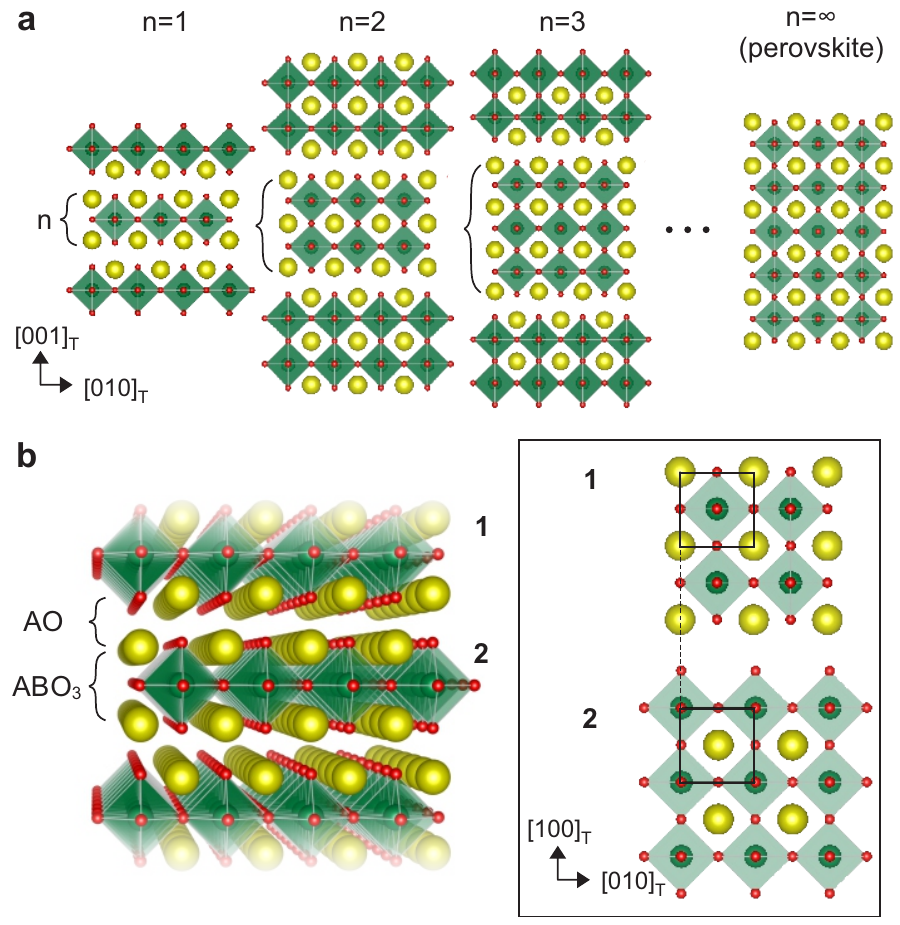}
    \caption{(a) Atomic models of RP oxides (A$_{n+1}$B$_{n}$O$_{3n+1}$). (b) Illustration showing the AO rock-salt layer and perovskite slabs in RP oxides. Alignment between neighboring perovskite slabs \textbf{1} and \textbf{2} is viewed from the out-of-plane ($[001]_T$) direction in a box. The subscript $T$ is tetragonal. The conventional unit cell is indicated with solid lines.}
    \label{fig:Sr2SnO4}
\end{figure}

In addition to structural distortions and local defects, many ABO$_3$ perovskites also form extended defects such as 2-dimensional stacking faults to accommodate nonstoichiometry \cite{Tilley2008}. Furthermore, double-AO stacking faults are often observed to order as well and give rise to the so-called Ruddlesden-Popper (RP) structures with the chemical formula A$_{n+1}$B$_{n}$O$_{3n+1}$ \cite{Tilley1977}. 
The n$^{th}$ member of the RP series consists of n-unit-cell-thick perovskite slabs with an additional AO rock-salt layer between the slabs \cite{Ruddlesden1957, Ruddlesden1958} (Fig.~\ref{fig:Sr2SnO4}a). The network of corner-sharing BO$_6$ octahedra in the perovskite structure is broken by the extra AO layer, and hence the structure consists of relatively isolated 2-dimensional slabs stacked on top of each other in a body-centered tetragonal fashion. As a result, various material properties can be tuned by the thickness of the perovskite slabs, as $n$ goes from $n=1$ in A$_2$BO$_4$ to infinity in ABO$_3$ \cite{Moon2008,Mulder2013,Birol2011}. As in the ABO$_3$ perovskites, the perovskite slabs in RP oxides exhibit various structural distortions from their high-symmetry structure including octahedral tiltings, octahedral distortions, and cation displacements that may give rise to ferroelectricity \cite{Woodward1997a,Balachandran2013,Li2020,Lee2013}. In addition, when subjected to external stimuli such as temperature, pressure, or strain, RP oxides can undergo even more complex structural modifications, which subsequently affects materials' properties \cite{Ramkumar2021,Lu2016}.

In this article, we present a comprehensive study of the crystal structure of the $n=1$ RP Sr$_2$SnO$_4$, which is a cousin of the perovskite SrSnO$_3$. SrSnO$_3$ has been actively studied as an ultra-wide bandgap material and has a nontrivial structural phase diagram as a function of temperature and strain \cite{Glerup2005,Zhang2017,Wang2018,Prakash2021,Kim2022}. 
Sr$_2$SnO$_4$ is one of two experimentally synthesized RPs from SrSnO$_3$, ($n=1,2$ Sr$_{n+1}$Sn$_n$O$_{3n+1}$)\cite{Green1996,Green2000} and is positioned at the end of the RP series (farthest from SrSnO$_3$, Fig.~\ref{fig:Sr2SnO4}a), thus, understanding its properties is essential to comprehend characteristics of RP series in comparison with perovskite. Sr$_2$SnO$_4$ also has a wide bandgap and, when doped, can be utilized as a phosphor due to its luminescence properties such as long-lasting phosphorescence \cite{Ueda2006,Kamimura2014}, mechanoluminescence \cite{Kamimura2012}, and thermoluminescence \cite{Wang2020}. Ceramics of Sr$_2$SnO$_4$ are synthesized with various methods, and optical/dielectric characterizations have been performed by multiple groups \cite{Kumar2020,Nirala2020,Kumar2018}. Yet, interestingly, the crystalline structure of Sr$_2$SnO$_4$, the most fundamental characteristic of the material, is still disputed. Neutron diffraction experiments performed by Green et al. point to a tetragonal structure with a space group of $P4_2/ncm$ (\#138) at $T = 12$~K, and an orthorhombic $Bmab$ (\#64) phase at $T = 295$~K \cite{Green1996}. On the other hand, Fu et al. report that Sr$_2$SnO$_4$ has an orthorhombic $Pccn$ (\#56) structure at $T = 4$~K and undergoes phase transitions of $Pccn\rightarrow Bmab\rightarrow I4/mmm$ (\#139) phases with increasing temperature \cite{Fu2002,Fu2004}. The reported structures have different SnO$_6$ oxygen octahedral rotation patterns in addition to rotation angles. While some of these patterns and the resulting space groups, such as $P4_2/ncm$, are also reported in other $n=1$ RP compounds, Sr$_2$SnO$_4$ is the only stoichiometric RP for which the space group  $Pccn$ is reported to our best knowledge \cite{Balachandran2013}. Additional disagreements exist in the reported structural transition temperatures, which might imply that the structure is highly sensitive to extrinsic effects \cite{Kumar2018}.

In order to resolve the disagreements in the experimentally reported crystal structures and assess the possible sensitivity of optical and electronic properties to boundary conditions such as strain, we perform a first-principles and symmetry-based analysis of Sr$_2$SnO$_4$. The first-principles density functional theory (DFT), in combination with the standard local density approximation or generalized gradient approximations, is expected to accurately predict the crystal structure and lattice energetics of this compound, since it is a band insulator with a rather simple electronic structure. The symmetry analysis is based on application of group theory, which involves labeling possible ionic displacements (i.e. normal modes) in crystals by the irreducible representations (irreps) of the space groups and is commonly used to describe the octahedral rotations in perovskites and related systems \cite{Balachandran2013_2,Stokes2002}. For example, out-of-phase and in-phase octahedral rotation patterns in perovskite oxides can be described as atomic displacements associated with the three-dimensional irreps $R_4^+$ and $M_3^+$ of the cubic perovskite structure with space group $Pm\bar{3}m$ (\#221) \cite{Balachandran2013_2}. Likewise, the octahedral rotation pattern in the $n=1$ RP Sr$_2$SnO$_4$ structure can be depicted as lattice distortions mediated by irreps of the high-symmetry body-centered tetragonal $I4/mmm$ (\#139) structure. 
We identify the unstable structural modes (imaginary frequency phonon modes) in the $I4/mmm$ structure by calculating the phonon dispersions, and then determine the irreps that they transform as accordingly. This enables finding possible meta-stable low-symmetry structures using group theory and predicting their energies by relaxing the crystal structures using DFT. 

Next, we explore the biaxial strain and hydrostatic pressure effects on the structure and symmetry also by employing the group-theoretical analysis and DFT. Biaxial strain, imposed on thin films by lattice mismatched substrates, can lead to significant changes in the crystal and electronic structure of crystalline materials \cite{Schlom2007, Schlom2014, Wang2018}. In addition to changing the crystal structure, biaxial strain may also have a significant effect on the electronic properties of materials. For example, the bandgap of a large number of oxide and halide perovskites are predicted to be strongly strain dependent \cite{Berger2011, Grote2015, Teply2021}, strain induced magnetic and coupled metal-insulator transitions coexist in SrMoO$_3$ \cite{Callori2015, Lee2011}, and the correlation strength in correlated metallic SrVO$_3$ and CaVO$_3$ are found to be strain tunable through the changes in the crystal structure \cite{Paul2019Vanadate}. The stannate RPs are wide bandgap semiconductors, and hence an electronic phase transition is not likely under strain, but changes in their bandgap and effective mass can nevertheless be significant enough to affect their optical and electronic properties. Hence, we study the impact of the strain-driven structural changes on the electronic properties by calculating the electronic band structures of Sr$_2$SnO$_4$ with different symmetry crystal structures. Lastly, the effect of hydrostatic pressure on the Sr$_2$SnO$_4$ structure is investigated and compared with the biaxial strain effects.

\section{Methods}

DFT calculations were performed by using the Vienna ab-initio simulation package (VASP) \cite{Kresse1993,Kresse1999} with the projector-augmented wave approach \cite{Blochl1994} and Perdew-Burke-Ernzerhof (PBE)-sol functional \cite{Perdew2008}. $10^{-6}$~eV was used as the convergence threshold in the self-consistent field cycle for the electronic wavefunctions. 
The energy cut-off for the plane wave basis set was set to 550~eV. The Brillouin zone of the tetragonal $I4/mmm$ structure ($4.068\times 4.068\times 12.560$ \AA) was sampled with a $12\times 12\times 6$ Monkhorst-Pack k-point grid, which was scaled down accordingly for larger supercells used for distorted crystal structures. As the convergence criterion for  ionic relaxations, a force per atom of $5\times 10^{-4}$~eV/\AA ~was used. The energy and volume of optimized structures are presented per formula unit (f.u.) of Sr$_2$SnO$_4$. Phonon calculations were carried out from a $2\times 2\times 2$ primitive supercell of the $I4/mmm$ structure of Sr$_2$SnO$_4$ using the frozen phonon approach. Dispersion curves were computed with the Phonopy software (v2.9.1) interfaced with VASP \cite{Togo2015}. Non-analytical term correction was incorporated with dielectric tensor and Born effective charges calculated using density functional perturbation theory as implemented in VASP \cite{Baroni2001}. The atomic structures corresponding to individual modes were constructed using the eigen-displacements, and irreducible representations (irreps) of each mode were determined by using ISODISTORT \cite{Campbell2006}. Construction of distorted structures adapting normal modes with different order parameters and directions was also conducted using ISODISTORT. The Bilbao Crystallographic Server was used as a reference for group theory tables \cite{Aroyo2014}. Crystal structure visualization was carried out using the VESTA software \cite{Momma2011}. Simulation data are available at the Data Repository for the University of Minnesota \cite{Yun2022}.

\section{Results and Discussion}

\subsection{Crystal structure of bulk Sr$_2$SnO$_4$}

\subsubsection{Experimentally Observed Structures} 

At high temperature, Sr$_2$SnO$_4$ has the high-symmetry reference RP structure with the space group $I4/mmm$ (\#139) \cite{Fu2004}. In this structure, there are isolated perovskite slabs that are single-octahedron-thick, and are half-unit-cell shifted along the $[110]_T$ direction with respect to their neighbors, as shown in Fig.~\ref{fig:Sr2SnO4}b. (Throughout this manuscript, all directions and planes are indexed using the axes of the conventional unit cell of the high-symmetry body-centered tetragonal structure.) All of the experimentally reported crystal structures at room temperature have structural distortions that break translational symmetry of the $I4/mmm$ RP structure and give rise to new Bragg peaks at the $X$ point of the Brillouin zone\cite{Green1996,Fu2002,Fu2004,Graef2012}. 
These distortions are due to oxygen octahedral rotations (or tilts) which are commonly observed and are classified in detail in multiple $n=1$ RP oxides \cite{Hatch1987, Hatch1989}. The three experimentally observed space groups ($Pccn$, $Bmab$, and $P4_2/ncm$) are all subgroups of the high-symmetry $I4/mmm$ space group of the RP structure, which itself is observed above 573~K \cite{Fu2004, Fu2002, Green1996}. These three structures can be obtained by the same octahedral rotation mode, which transforms as the $X_3^+$ irrep, and differ only in the direction of 2-dimensional order parameter: $X_3^+(a;a)$, $X_3^+(a;0)$, and $X_3^+(a;b)$ reduce the symmetry to $P4_2/ncm$, $Bmab$ (or equivalently $Cmca$), and $Pccn$ respectively. The tetragonal space group $P4_2/ncm$ ($X_3^+(a;a)$) has out-of-phase oxygen octahedral tilts only along a single in-plane Cartesian axis, but this direction is alternating on neighboring perovskites slabs. In other words, it corresponds to an alternating $a^-b^0c^0$ and $a^0b^-c^0$ rotation pattern in the Glazer notation. The base-centered orthorhombic $Bmab$ ($X_3^+(a;0)$) is obtained by out-of-phase tilts along a $[110]_T$ direction, which is $a^-a^-c^0$ in the Glazer notation. The experimentally observed $Pccn$ structure, on the other hand, involves rotations around low-symmetry directions on the $a$-$b$ plane. The atomic model is illustrated in Supplemental Material (SM) Fig. S1 \cite{SM}.

\begin{figure*}
    \centering
    \includegraphics[width=0.8\textwidth]{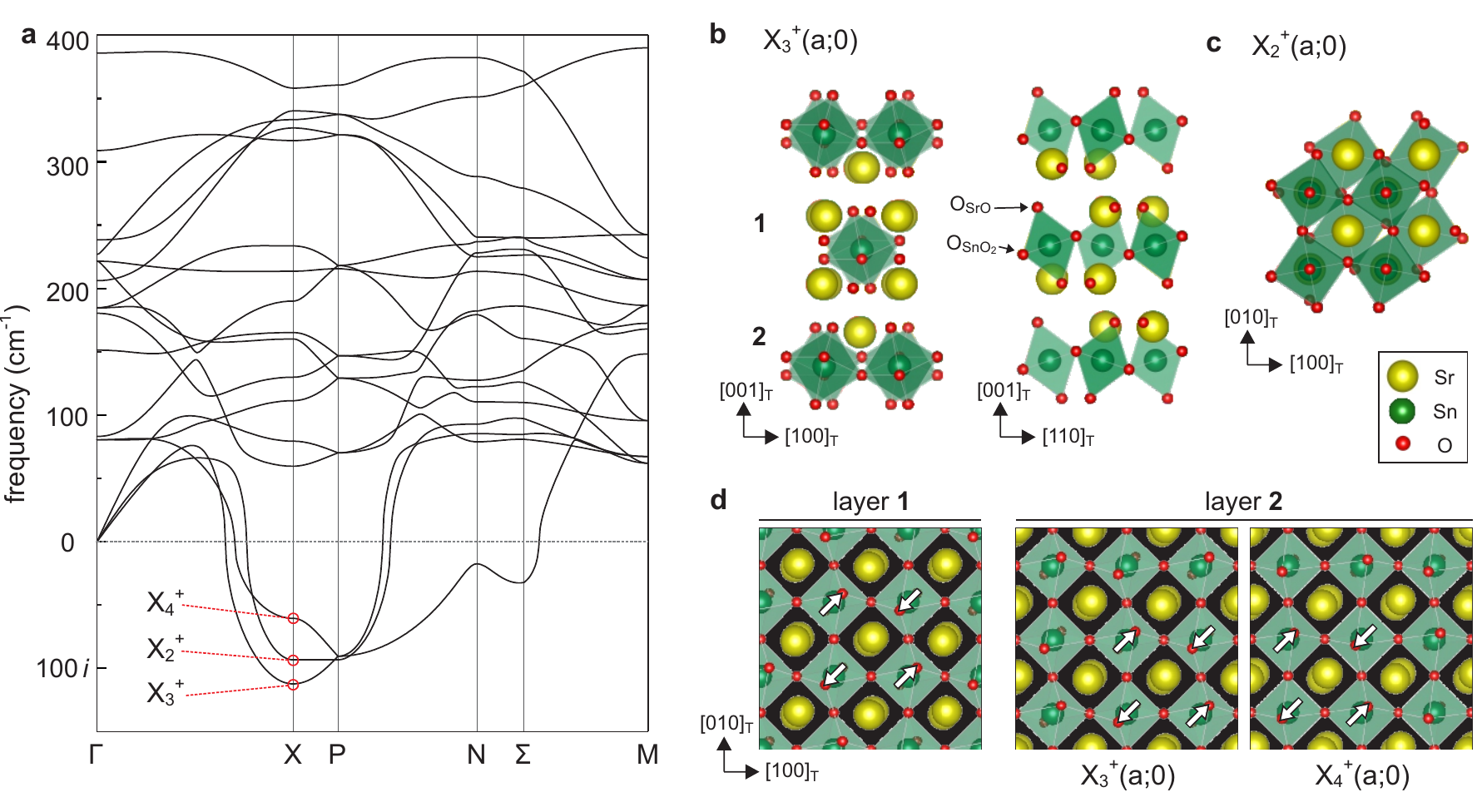}
    \caption{(a) Phonon dispersion diagram of $I4/mmm$ Sr$_2$SnO$_4$. Unstable $X$ modes with imaginary frequencies are marked. The band path is visualized in SM Fig. S2. (b,c,d) Atomic models of Sr$_2$SnO$_4$ structure transformed with the three unstable $X$ modes. (b) $X_3^+(a;0)$ mode-implemented structure exhibiting OOT. (c) $X_2^+$ mode-implemented structure showing OOR. (d) Atomic models describing different alignments between neighboring perovskite slabs in the $X_3^+$ and $X_4^+$ irreps-associated structures. When the structures are viewed from the out-of-plane ($[001]_T$) direction, and the layer \textbf{1} is fixed, the configurations of the layer \textbf{2} in $X_3^+$ and $X_4^+$ irreps are dissimilar. The direction of the atomic displacements of $O_{SrO}$ due to OOT are indicated with the arrows.}
    \label{fig:Phonon_dispersion_structure}
\end{figure*}

\subsubsection{Lattice dynamics of Sr$_2$SnO$_4$}

To investigate the zero temperature ground-state (or the lowest energy) crystal structure of Sr$_2$SnO$_4$, we perform first-principles lattice dynamics calculations in the $I4/mmm$ structure and identify lattice instabilities that may freeze in and reduce the symmetry, as well as the energy. The phonon dispersion is shown in Fig.~\ref{fig:Phonon_dispersion_structure}a. Phonons with imaginary frequencies, which indicate lattice instabilities, are present at the $X$, $P$, and $N$ k-points of the Brillouin zone. 
Since all the structures proposed based on experimental data are associated with the $X_3^+$ irrep, we focus our analysis on the unstable modes at the $X$-point only, but consider all three unstable $X$-modes. These modes break the translational symmetry in the same way, and hence lead to primary Bragg spots at the same positions but with different intensities in diffraction experiments. While there are different systematic absences due to nonsymmorphic symmetries in these space groups, some of the allowed peaks may be hard to detect due to the low structure factor of oxygen atoms. As a result, differentiating the different structures the $X$ modes lead to might be experimentally challenging, and thus all of them need to be considered to theoretically predict the ground state structure. In the SM section 3, we also provide a partial analysis of the unstable $N$ and $P$-modes. This analysis shows that they lead to monoclinic structures without larger energy gains than the $X$-modes, and therefore, are not critical for the present discussion.

We identified the three unstable modes at the $X$-point with the $X_3^+$, $X_2^+$, and $X_4^+$ irreps of the $I4/mmm$ space group. All of these irreps correspond to SnO$_6$ octahedral rotations about different crystallographic axes as well as associated Sr displacements. The $X_3^+$ and $X_4^+$ irreps both correspond to out-of-phase oxygen octahedral tilts (OOT) about the in-plane axes ($[100]_T$ and $[010]_T$) but with different phases along the out-of-plane direction, while the $X_2^+$ irrep corresponds to SnO$_6$ oxygen octahedral rotations (OOR) about the out-of-plane ($[001]_T$) axis. These modes are illustrated in Fig.~\ref{fig:Phonon_dispersion_structure}b-c. All three $X$ irreps are 2-dimensional, and hence have three possible order parameter directions (OPD)($(a;0)$, $(a;a)$, and $(a;b)$) that lead to different space groups.

The oxygen octahedral rotation modes in RP's are different from their counterparts in bulk perovskites in two important aspects. The first difference is that while the octahedral rotations can induce (improper) displacements of A-site cations in perovskites, there is no coupling between rotations and A-site displacements at the harmonic ($2^{nd}$) order, and hence, the octahedral rotation phonon eigenvectors of perovskites don't include any A-site displacements. However, in RP's, the phonon eigenvectors include small but nonzero A-site displacements as well. For example, the $X_3^+(a;0)$ mode includes atomic displacements of Sr in addition to O$_{SrO}$ (apical O in the SrO$(001)$ layer), and O$_{SnO_2}$ (O in the SnO$_2(001)$ layer). 
The other difference is due to the broken connectivity of the octahedral network and the shift of the neighboring perovskite layers. Unlike in perovskites, where there is only one symmetry allowed way to have an in-phase (out-of-phase) octahedral rotation along a certain axis, there is more than one degree of freedom in the RP's that correspond to an in-phase (out-of-phase) octahedral rotation along an axis. 
For example, both the $X_3^+$ and $X_4^+$ irreps correspond to out-of-phase OOT around in-plane axes. However, the alignment between neighboring perovskite slabs of the $X_4^+$ displacement pattern is different from that of the $X_3^+$ mode, which accounts for their disparate symmetries and stabilities. When the relative in-plane shift between the neighboring perovskite slabs, e.g. layers \textbf{1} and \textbf{2} in Fig.~\ref{fig:Phonon_dispersion_structure}b and d, is parallel to the octahedral rotation axis, i.e. perpendicular to the in-plane displacement direction of O$_{SrO}$  (white arrows in Fig.~\ref{fig:Phonon_dispersion_structure}d), a structure with a $X_3^+(a;0)$ mode is produced, and when the shift is perpendicular to the rotation axis, i.e. parallel to the O$_{SrO}$ displacements, a $X_4^+(a;0)$ mode is generated. See SM section 4 for details. 
Despite their similarities, the $X_3^+$ mode is more unstable than the $X_4^+$ mode as can be seen in Fig.~\ref{fig:Phonon_dispersion_structure}a, implying that a structure incorporating the $X_3^+$ distortion pattern reduces the structural energy more at quadratic order. We address why this may be from a bond-valence point of view in the following sections.

\subsubsection{Order parameter directions}
In the previous section, we compared the relative strengths of lattice instabilities. All three of these instabilities are 2-dimensional, and higher order terms in the Landau free energy leads to different energy gains for different OPDs. 
In this subsection, we consider these different order paramater directions to map out the energy surface, and predict the lowest energy structure. For each of the three 2-dimensional irreps, there are three distinct phases, corresponding to different values of OPDs, $a$ and $b$. For example, as discussed earlier, freezing in the $X_3^+(a;0)$ mode in the $I4/mmm$ structure leads to a $Bmab$ phase with OOT around the $[110]_T$ axis, i.e. equal OOT around $[100]_T$ and $[010]_T$ axes. In contrast, an $X_3^+(a;b)$ distortion leads to a $Pccn$ phase exhibiting OOT with dissimilar tilts around $[100]_T$ and $[010]_T$, and an $X_3^+(a;a)$ leads to the tetragonal $P4_2/ncm$ where the octahedral rotation axes alternate between $[100]_T$ and $[010]_T$ in consecutive perovskite layers. (These distortions are illustrated in detail in SM Fig. S5.)
 Similarly, there are three phases each for the $X_2^+$ and $X_4^+$ modes, corresponding to distinct OPDs, which are shown in SM Fig. S6. 

\begin{figure}
    \centering
    \includegraphics[width=0.9\columnwidth]{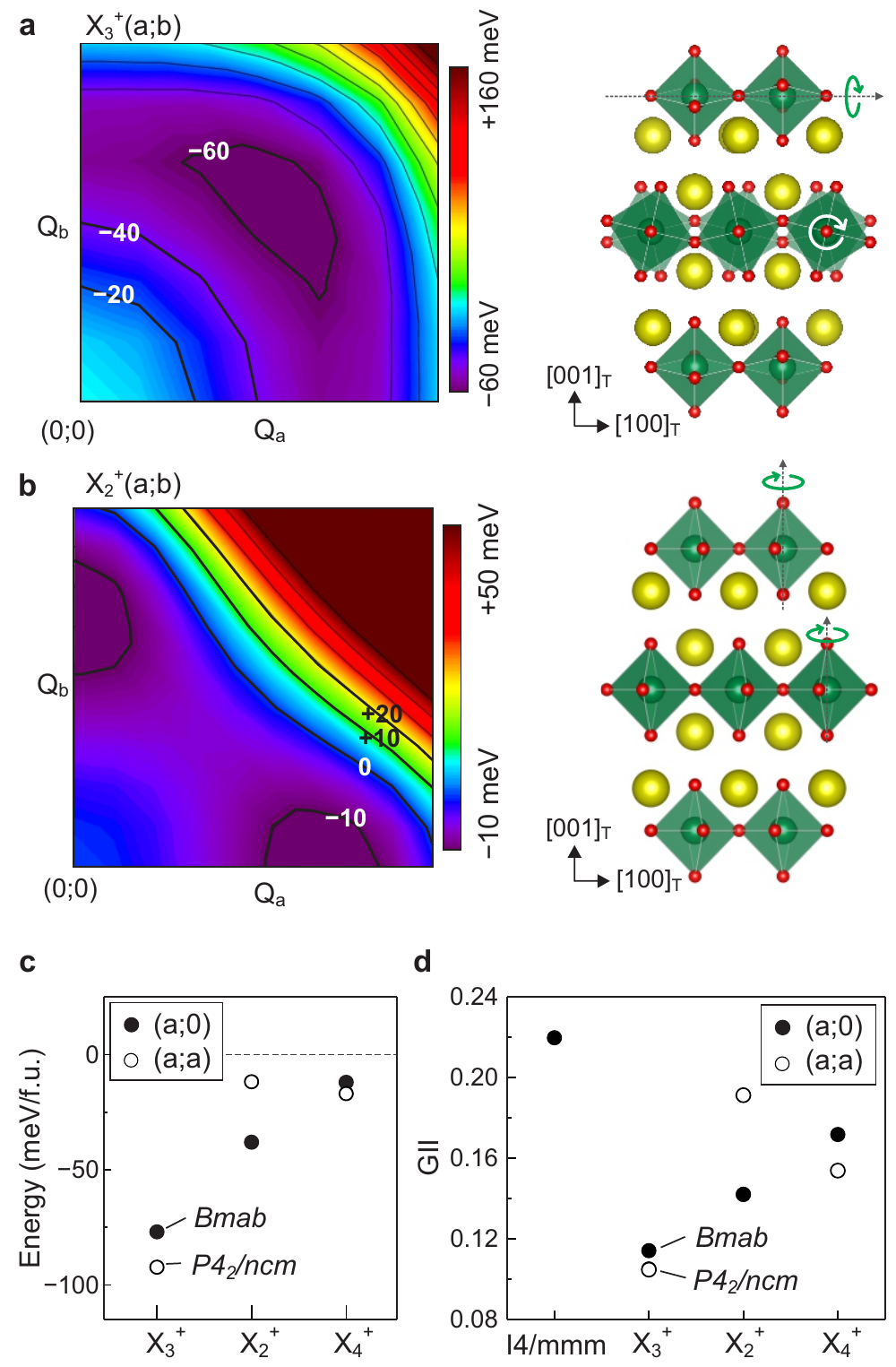}
    \caption{(a,b) Three-dimensional (3D) energy map of Sr$_2$SnO$_4$ structures containing $X(a;b)$ irreps as a function of mode amplitudes of a and b. 3D maps were constructed for the $X_3^+$ (a) and $X_2^+$ (b) irreps. The minimum energy structure was further relaxed, and the optimized structures are presented on the right. (c) The energy and (d) GII values of the relaxed structures with different irrep modes.}
    \label{fig:E3Dmap}
\end{figure}

 To optimize structures for each $X$ mode, we calculate the energy of structures with different $(a;b)$ values on a grid for different irreps from DFT. Fig.~\ref{fig:E3Dmap}a shows the energy map of a Sr$_2$SnO$_4$ structure distorted with an $X_3^+$ mode as a function of amplitudes (Q) along the two order parameter components. The phonon eigenvectors are used to determine the displacement patterns for each order parameter component; in other words, the ratio between the three types of atomic displacements (Q$_{Sr}$:Q$_{O_{SrO}}$:Q$_{O_{SnO_2}}$) is fixed to the values obtained from phonon calculation. 
We find that for $X_3^+$, the minimum energy is obtained for $a=b$ (space group $P4_2/ncm$) whereas for $X_2^+$, the energy is lowest when only one of the two order parameter components is nonzero (space group $Bmab$). For $X_4^+$ (SM Fig. S10), the minimum energy is also obtained for $a=b$ (space group $P4_2/nnm$). Of all these structures, the lowest energy is obtained for $X_3^+(a;a)$, and thus, the lowest energy structure in the absence of strain is $P4_2/ncm$. 

In passing, we note that the fact that we find local minima only for either \textit{(i)} when $a=b$, or \textit{(ii)} when one of $a$, $b$ is zero, is consistent with a 6$^{th}$ order Landau free energy expansion. Symmetry imposes strict conditions on the allowed terms in the expansion of the free energy as a polynomial of the order parameter components around the high-symmetry structure, which is the undistorted $I4/mmm$ RP structure in this case \cite{Hatch2003}. All three unstable $X$-point irreps $X_i^+$ ($i=1,~2,~3$) of Sr$_2$SnO$_4$ have the same trivial Landau Free energy up to 6$^{th}$ order: 
\begin{multline}
	\mathcal{F}(a,b)=\alpha(a^2+b^2)+\beta_4(a^2+b^2)^2+\gamma_4(a^2b^2)
	\\+\beta_6 (a^2+b^2)^3 +\gamma_6 (a^2+b^2)(a^2b^2),
	\label{equ:landau}
\end{multline}
where the Greek letters are material specific coefficients that can be obtained from DFT. The $(a^2+b^2)$ terms do not depend on the relative direction of the order parameter, and hence cannot differentiate between $X_i^+(a;0)$, $X_i^+(a;b)$, and $X_i^+(a;a)$. The only terms that depend on the direction of the order parameters are the $\gamma$ terms, which have the same type of $\sim a^2 b^2$ dependence. 
Depending on the signs of $\gamma$, this expression is minimized either for $|a|=|b|\neq 0$ or for one of the two components being zero. Thus, the space group $Pccn$, which is obtained by the $X_3^+(a;b)$ OPD, requires higher order terms in the free energy expansion to be important in determining the ground state crystal structure. Our analysis (not shown) of the higher order Landau free energy expansion for $X_3^+$ shows that the 8$^{th}$ order terms indeed allow solutions with $|a|\neq|b|\neq 0$. Since in most crystalline materials a lattice free energy expansion up to just 6$^{th}$ order is sufficient to determine the correct low temperature crystal structure, we find that the space group $Pccn$ is an unlikely ground state candidate for Sr$_2$SnO$_4$. 

In principle, both strain-coupling and higher order terms can be important in determining the symmetry of the lowest energy structure. In order to take these effects into account completely, we perform relaxations of the crystal structure of Sr$_2$SnO$_4$ using DFT, and allowing both the internal atomic coordinates and the lattice vectors to relax. Since the DFT calculations preserve the symmetry and the basin of attraction of different metastable phases can be large, for each OPD we use multiple starting points with $(a;0)$, $(a;a)$, and $(a;b)$. Note that these complete ionic relaxations allow the Q$_{Sr}$:Q$_{O_{SrO}}$:Q$_{O_{SnO_2}}$ ratio, as well as the strain state of the lattice, to change. We could not stabilize a state with $a\neq b$ in any of these calculations. All calculations initiated with such a structure relax to a higher symmetry structure with either $(a;a)$ or $(a;0)$. While they do not make a qualitative difference in the ground state, we find the strain effects to be quantitatively important: For example, when relaxed, the $P4_2/ncm$ phase has in-plane and out-of-plane lattice constants change by +0.83\% and -1.38\% compared to the undistorted $I4/mmm$ structure with the same supercell size. All in all, the lowest energy structure is still $P4_2/ncm$ obtained by $X_3^+(a;a)$, and the minimum energy for structures with $X_2^+$ and $X_4^+$ distortions are obtained at OPDs $X_2^+(a;0)$ and $X_4^+(a;a)$, respectively, as shown in Fig.~\ref{fig:E3Dmap}c.

While the $X_3^+(a;a)$ irrep gives the most stable structure, the other unstable modes may also be present in the lowest energy structure in combination with the $X_3^+$ mode. In order to check this possibility, we also relaxed structures that have a superposition of $X_3^+$ and the $X_2^+$ or $X_4^+$ modes. These structures all relax to a structure containing only the $X_3^+(a;a)$ mode. As a further check, we also performed structural relaxations of the experimentally-determined low-temperature crystal structure with the space group $Pccn$ and the $X_3^+(a;b)$ distortion \cite{Fu2002}. This structure also relaxed to the same final state with the $X_3^+(a;a)$ distortion. As a result, at the DFT level, there is no reason to doubt that the lowest energy crystal structure of Sr$_2$SnO$_4$ is obtained by an $X_3^+(a;a)$ distortion of the high-symmetry $I4/mmm$ structure and has the space group $P4_2/ncm$.

\subsubsection{Bond valence approach}
To gain further insight into the structural stability of the different phases in Sr$_2$SnO$_4$, we use the bond valence approach. The bond valence sum and the related quantity of the Global Instability Index (GII) have been proven to be effective tools to explain structural stability of crystals where modification of the local bonding environment can explain structural transitions \cite{Lufaso2001,Brown2009,Brown1978,Salinas-Sanchez1992, Brese1991}. As an example, different rotation patterns in perovskites were shown to be predicted by the optimization of bond valences more than two decades ago \cite{Woodward1997b}. 
The valence $V_i$ of an atom $i$ can be estimated by summing the individual bond valences $v_{ij}$ that are calculated from the bond length between the atoms:
\begin{equation}
	V_i=\sum_{j=1} v_{ij}
\end{equation}
and 
\begin{equation}
	v_{ij}=\mathrm{exp}((R_0-R_{ij})/b),
\end{equation}
where $R_0$ is the empirical value of the expected bond distance, $R_{ij}$ is the actual bond distance between atoms $i$ and $j$, and $b=0.37$~\AA ~is an empirical constant \cite{Brese1991, Brown2009}. The observed valence of an atom $i$ typically deviates from its formal valence ($V_{formal,Sr}$ = 2, $V_{formal,Sn}$ = 4) and the discrepancy is referred to as a discrepancy factor, $d_i=V_{i}-V_{formal}$. The discrepancy factor is an indicator of whether an atom is under-($d_i < 0$) or over-($d_i > 0$)bonded. The root mean square of all the discrepancy factors in a structure,  
\begin{equation}
	GII=\sum_{i=1}^{N}\left(\frac{d_i^2}{N}\right)^{1/2}, 
\end{equation}
 is referred to the GII. A small GII is an indicator of overall structural stability. 

In Fig.~\ref{fig:E3Dmap}d, we show the GII values for the DFT-optimized structures of Sr$_2$SnO$_4$. Whereas the original form of GII includes bond valences of all ions in an examined structure, our careful analysis shows that, interestingly, the stability of Sr$_2$SnO$_4$ is better-explained by GII using only cations; and thus, we report cation-based GII. (See SM section 7 for the GII calculation analysis.) The GII values of different phases follow the same trend as the DFT energies, which indicates that the phase stability of Sr$_2$SnO$_4$ can be predicted from the local bonding environment of cations. The bond valences of Sr (shown in SM Fig. S7) make the dominant contribution to the GII and determine the low-symmetry structure. This is in-line with the observation in perovskites that the octahedral rotations are A-site driven instabilities, and serve to optimize the bonding environment of the underbonded A-site cations.

\subsection{Strain effects on the ground-state structure of Sr$_2$SnO$_4$}

In this section, we evaluate the effect of biaxial strain on Sr$_2$SnO$_4$. We limit our discussion to (001) grown thin films (i.e. structures where the crystallographic $c$ axis is relaxed but the $a$ and $b$ lattice parameters are fixed by a substrate), since this is the most commonly grown orientation by, for example, molecular beam epitaxy method. (See, for example, \cite{Haeni2001, Lee2013, Burganov2016, Wang2018}.) 

\begin{figure}
    \centering
    \includegraphics[width=0.9\columnwidth]{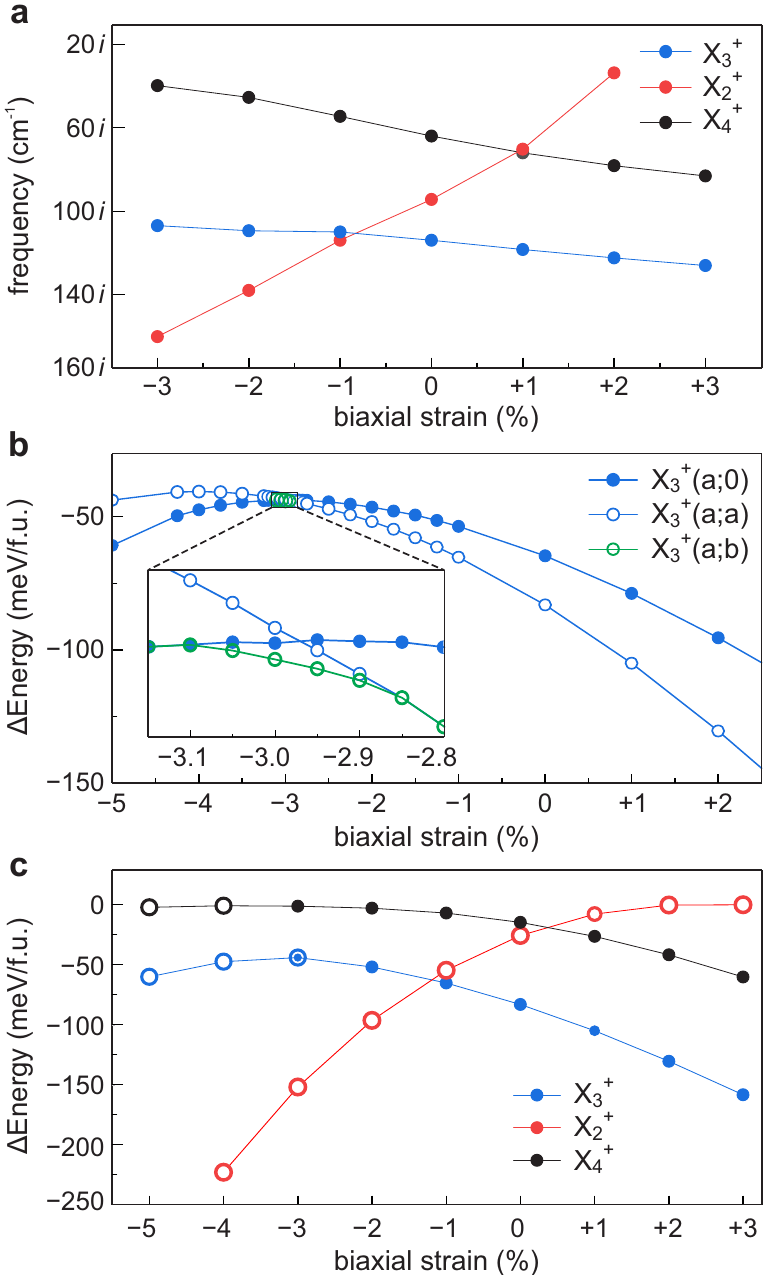}
    \caption{Strain effect on the $X$ modes. (a) Phonon frequency of the three $X$ modes. (b) The relative energy of the $X_3^+$ modes with dissimilar OPDs with respect to the $I4/mmm$ structure. The region where $X_3^+(a;b)$ is stabilized is magnified in the inset. (c) The relative energy of the optimized $X$ modes. OPD of each structure is indicated via the symbol types: closed circle-$X(a;0)$, open circle-$X(a;a)$, open circle with a center dot-$(a;b)$.}
    \label{fig:strain}
\end{figure}

As hinted from the distinct lattice parameter change upon relaxation of the minimum energy structures in the energy maps, each irrep mode responds to biaxial strain differently. In order to predict how the lattice instabilities evolve, we relax the high-symmetry $I4/mmm$ structure with in-plane biaxial strain ranging from -3 \% to +3 \% and calculate the phonon dispersions for each of these strained structures. We define strain for a structure with in-plane lattice parameter $a$ as $\eta=(a-a_0)/a_0$, where $a_0$ is the lattice parameter of the fully relaxed but undistorted ($I4/mmm$) structures. The phonon dispersion curves presented in SM Fig. S9 indicate not only that the 3 unstable $X$ modes remain unstable under strain, but also that the strongest instability remains to be at the $X$ point. 
The frequencies of the three relevant $X$-modes are shown in Fig.~\ref{fig:strain}a. Under compressive strain, the $X_3^+$ and $X_4^+$ instabilities (related to OOT around the in-plane axes) weaken while the $X_2^+$ becomes (associated with the out-of-plane axis OOR) even more unstable. On the other hand, under tensile strain, the $X_3^+$ and $X_4^+$ modes become slightly more unstable, while the $X_2^+$ instability is suppressed, and its frequency becomes real at +3 \%.

In Fig.~\ref{fig:strain}b, we show the energy of structures with $X_3^+$ distortions. In SM Fig. S10, 3D energy maps of the Sr$_2$SnO$_4$ structures at different strain degree are also presented. Extending the compressive strain range to beyond -3\% shows that there is a transition from $X_3^+(a;a)$ ($P4_2/ncm$) to $X_3^+(a;0)$ ($Bmab$). This transition can be understood to stem from a strain-induced sign change of the sum of the $\beta_2$ term in Eq.~\ref{equ:landau} with the higher order terms with the same dependence on $|a|/|b|$. As this term crosses zero, higher order terms that favor different $|a|/|b|$ ratios become apparently important, and the $X_3^+(a;b)$ ($Pccn$) phase has a lower energy in a narrow strain range as shown in the inset of Fig.~\ref{fig:strain}b. However, this does not make $Pccn$ experimentally observable either, since under compressive strain, the $X_2^+$ turns out to lead to a lower energy structure, as seen from Fig.~\ref{fig:strain}c where we present the DFT-predicted energies of structures obtained from each of the $X$ point instabilities. The transition from structures obtained by $X_3^+$ distortions under tensile strains to structures with $X_2^+$ octahedral rotations can be understood as a transition from a phase with octahedral tilts around in-plane axis to a phase with octahedral rotations around the out-of-plane axis. Such transitions are rather common in perovskite and perovskite-related structures and are primarily driven by the fact that out-of-plane octahedral rotations increase the in-plane B--O bond lengths, which becomes favorable under compressive strain. This enforces a shorter than ideal B--O distance. 

In passing, we note that the strain driven structural transitions observed in Fig.~\ref{fig:strain}c can also be predicted from the GII to a large extent, which means that electrostatics and local bonding are the determining factor for phase stability of Sr$_2$SnO$_4$ under strain as well. We show the details of strain dependent GII calculations in SM Fig. S11c-d.

\begin{figure}
    \centering
    \includegraphics[width=0.95\columnwidth]{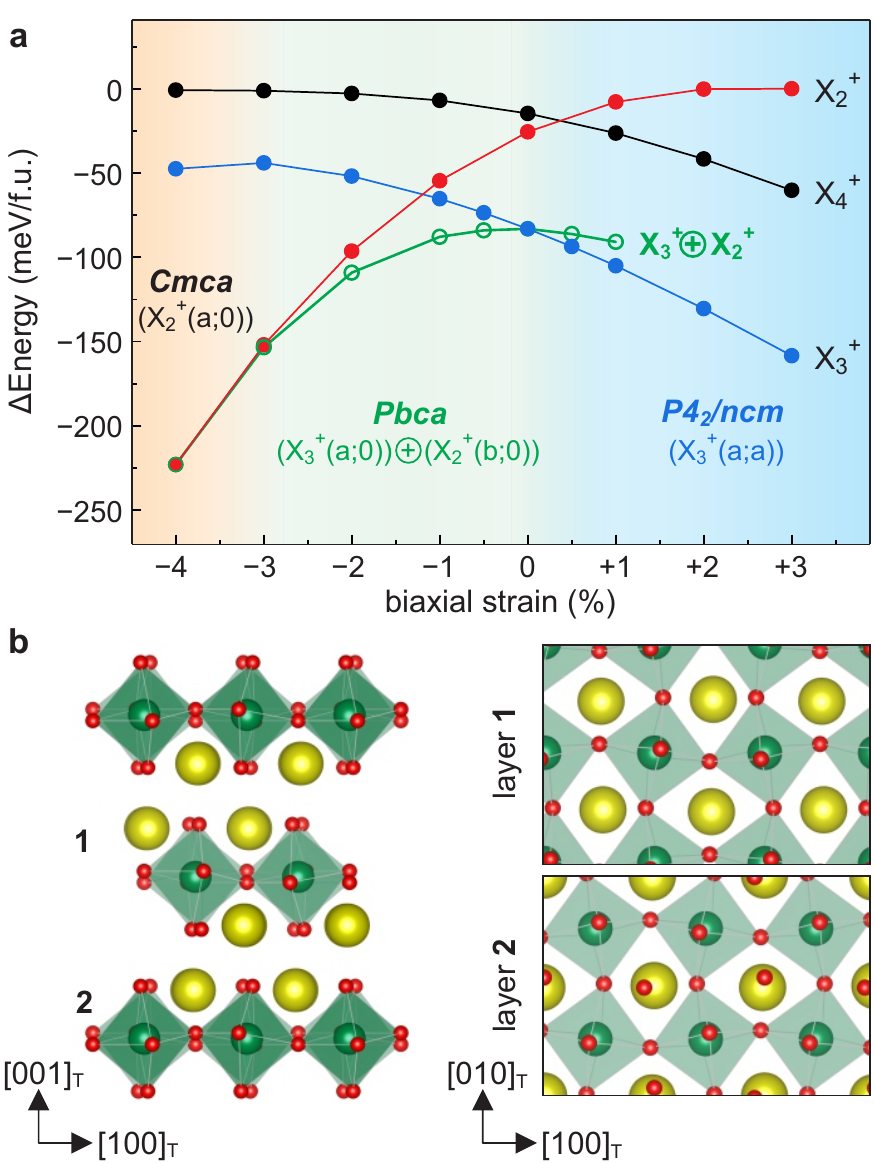}
	\caption{Combination of the $X$ modes under the biaxial strain. (a) The energy of structures mediated by the three $X$ irreps and by superposition of two irreps ($X_3^+ \oplus X_2^+$). (b) The atomic model of the optimized structures under the biaxial strain of -1\%. The structure is a $Pbca$ phase, and its distortion is described with the reducible representation of $X_3^+(a;0) \oplus X_2^+(b;0)$.}
    \label{fig:superposition}
\end{figure}

We finally consider the combinations of different $X$ point distortions to examine if any phases which is not considered in Fig.~\ref{fig:strain}c are stabilized by strain. The resultant phase diagram, Fig.~\ref{fig:superposition}a, hosts a $Pbca$ phase that has octahedral rotations around both in-plane and out-of-plane axis under moderate compressive strain. The $Pbca$ phase is obtained by a superposition of $X_3^+(a;0)$ and $X_2^+(b;0)$ distortions as illustrated in Fig.~\ref{fig:superposition}b. Therefore, our result highlights that as the phase transition from $P4_2/ncm$ to $Cmca$ takes place in Sr$_2$SnO$_4$ due to compressive biaxial strain, the intermediate phase $Pbca$ can be stabilized in the strain range of $-3\% \leq \eta \leq 0\%$. See SM Fig. S11 for stability comparison between the $Pbca$ and $P4_2/ncm$ (ground-state structure) phases as a function of strain.

\subsection{Strain effect on the electronic properties of Sr$_2$SnO$_4$}
\begin{figure}[h]
    \centering
    \includegraphics[width=1\linewidth]{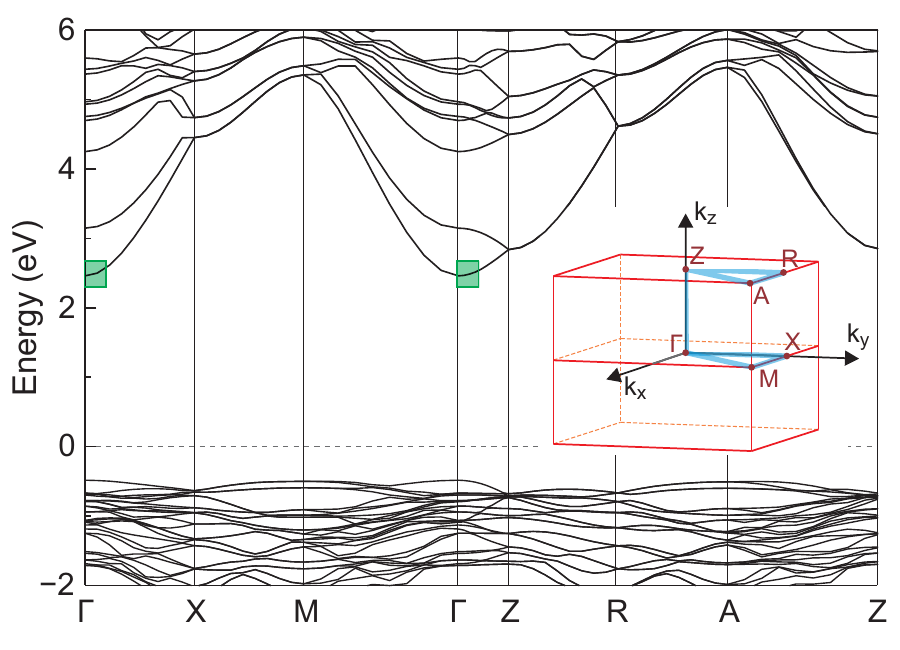}
    \caption{(a) The electronic band structure of the $P4_2/ncm$ (\#138) phase of Sr$_2$SnO$_4$ structure. The regions near the conduction band minimum (inside the green boxes) were used for the effective mass calculation. (inset) A schematic of the Brillouin zone of the $P4_2/ncm$. The band path used for the diagram is illustrated with the blue lines. }
    \label{fig:electronic_band_diagram}
\end{figure}

Perovskite stannates including SrSnO$_3$ are known to exhibit strain-sensitive electronic band structures, and strained thin films have been reported to have bandgaps larger by $\sim$1 eV than bulk \cite{Singh2014, Gao2020, Baniecki2017}. Given the similar character of valence and conduction bands in Sr$_2$SnO$_4$, similar changes that would affect its luminescence properties may be possible. The bottom of the conduction band of Sr$_2$SnO$_4$ is comprised of Sn $5s$ states, which hybridize with O $2p$ derived bands. Thus, the Sn-O bonding configurations such as Sn-O-Sn bonding angles controlled by octahedral rotations directly influence the bandgap and the electron effective mass. The electronic band diagram of the ground-state structure at zero strain, the $P4_2/ncm$ phase containing the $X_3^+(a;a)$ irrep, is shown in Fig.~\ref{fig:electronic_band_diagram}. 

\begin{figure}
    \centering
    \includegraphics[width=0.9\linewidth]{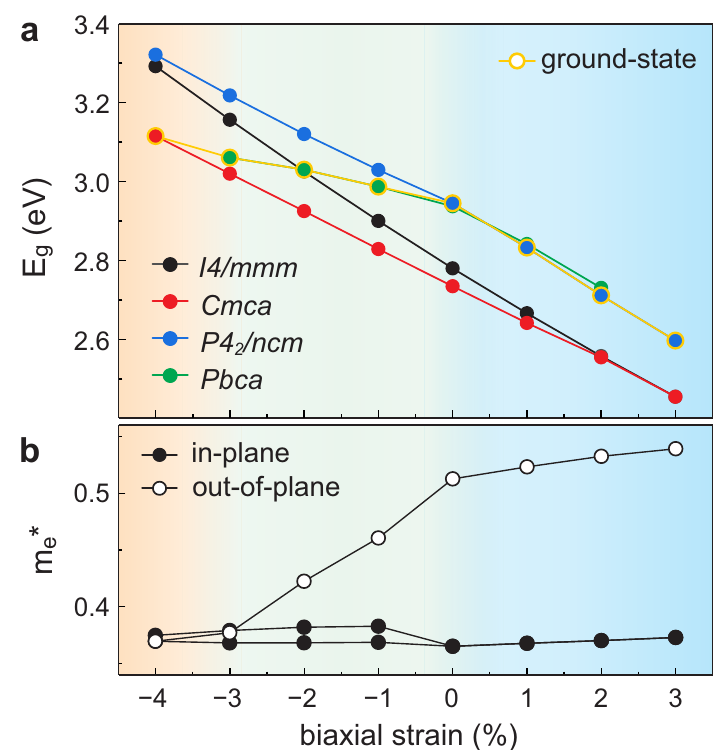}
	\caption{Electronic properties of Sr$_2$SnO$_4$ with respect to biaxial strain. (a) Bandgap. (b) Effective electron mass $m_e^*$ of ground-state structures along the in-plane and out-of-plane directions. Two in-plane effective masses, originating from the in-plane anisotropy in the orthorhombic phases, are plotted as closed circles.}
    \label{fig:electronic}
\end{figure}

The key feature of the conduction bands of perovskite stannates BaSnO$_3$ and SrSnO$_3$, a single parabolic s-band at the $\Gamma$ point, is also present in Sr$_2$SnO$_4$, and it is resistant to the structural distortions in all phases we considered. However, different degrees and patterns of octahedral rotations lead to different bandgaps, as shown in Fig.~\ref{fig:electronic}a. Following the lowest energy structure (yellow) as a function of strain, it is evident that larger in-plane lattice parameters reduce the bandgap by as much as $\sim$0.4~eV for 3\% strain. (Note that while the standard DFT+GGA calculations we perform underestimate the bandgap, they usually capture the correct trends with strain.) The same effect of tensile (compressive) biaxial strain decreasing (increasing) the bandgap was also observed in SrSnO$_3$ \cite{Gao2020} and suggests that strain engineering can be employed to achieve photoemission with different wavelengths from this material.

Next, we examine the electron effective mass of Sr$_2$SnO$_4$, which is an important quantity for perovskite stannates as high mobility n-type semiconductors \cite{Prakash2019, He2020, Kim2022}. In Fig.~\ref{fig:electronic}b, the electron effective mass calculated in the lowest energy crystal structure is presented as a function of biaxial strain. The symmetry of the Sr$_2$SnO$_4$ structures lead to dissimilar effective masses along the out-of-plane and the in-plane directions. (See SM section 12 for details.) Under zero strain, while the in-plane effective mass is similar to that of perovskite SrSnO$_3$ ($\sim$0.4m$_e$)\cite{Khuong2015,Wang2018}, the out-of-plane effective mass is higher by $\sim$30\%, which is in line with the broken Sn-O network in this direction. Interestingly, under large compressive strain, the effective mass tensor of the $Pbca$ and $Cmca$ phases become closer to isotropic, and the out-of-plane effective mass becomes comparable to the in-plane one. While compressive strain leads to both SnO$_6$ octahedral distortion, i.e. elongation along the c-axis, and tilting, i.e. OOT to OOR, the reduction of the electron effective mass indicates that the change of the octahedral tilting patterns plays a major role in determining the effective mass of Sr$_2$SnO$_4$. As OOT weakens under compressive strain, the out-of-plane direction O-Sn-O bonds are aligned to the c-axis, and consequently, the effective mass along the direction decreases. The isotropic and perovskite-comparable electron effective mass, in combination with the enhanced bandgap, under compressive strain renders Sr$_2$SnO$_4$ as a viable wide-bandgap semiconductor.

\subsection{Hydrostatic pressure effect}
\begin{figure}
    \centering
    \includegraphics[width=0.9\columnwidth]{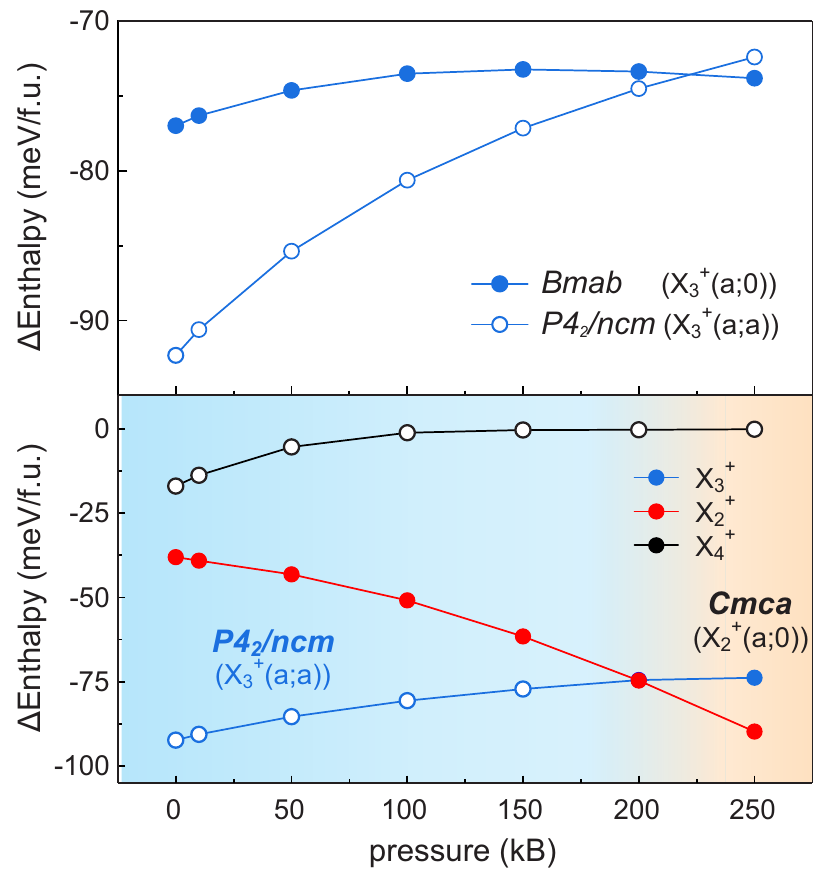}
	\caption{The effect of pressure on the $X$ modes of Sr$_2$SnO$_4$. (Top) Relative enthalpy of $X_3^+$ mode with different OPDs. (Bottom) Relative enthalpy of $X$ mode-adapted structure with respect to the $I4/mmm$ structure. The closed and open circles represent the order parameter configurations of $(a;0)$ and $(a;a)$, respectively. Space group of the minimum energy structure at different pressures is shown as the background color.}
    \label{fig:Pstress}
\end{figure}

Lastly, we consider the effect of hydrostatic pressure on the structural phases of Sr$_2$SnO$_4$. Different symmetry structures are relaxed under fixed stress boundary conditions, and the enthalpy $H$ ($H=E-PV$, where $E$ is total energy, $P$ is pressure, and $V$ is a volume) of the optimized structures are compared to examine their relative stability (see Fig.~\ref{fig:Pstress}). Pressures up to 250 kBar are considered. The pressure response of Sr$_2$SnO$_4$ is found to be very similar to its response to biaxial compressive strain, which also reduces the unit cell volume. 
With increasing pressure, the structure with $X_3^+(a;0)$ and $X_4^+(a;0)$ distortions become energetically less favorable over the structures with $X_3^+(a;a)$ and $X_4^+(a;a)$ modes. Until $P$= ~200 kBar, the structural ground state does not change, while above this pressure the $Cmca$ structure hosting $X_2^+(a;0)$ distortions becomes the lowest energy structure. 
We find that in contrast to biaxial strain, neither a combination of multiple structural distortions, nor a different direction of these three order parameters, are stabilized under pressure.

\section{Conclusions}

In this paper, we provide a comprehensive first-principles and group-theoretical study of the crystal and electronic structures of $n=1$ RP Sr$_2$SnO$_4$. Our main findings are that \textit{(i)} the lowest energy structure at ambient pressure is $P4_2/ncm$, and the next lowest energy metastable phase is $Bmab$, which is consistent with the experimental observation of a phase transition between these two phases, followed by a transition to the $I4/mmm$ phase with increasing temperature\cite{Green1996}, \textit{(ii)} the $Pccn$ phase is not even metastable (except for a narrow range under large biaxial strain), and therefore not likely to be present in a bulk material, \textit{(iii)} biaxial strain leads to both structural phase transitions and significant changes in the bandgap and electron effective mass, which is relevant to both semiconducting and optical properties of the material, \textit{(iv)} structural phase transition to $Cmca$ occurs above 200 kPa without necessarily producing the intermediate $Pbca$ phase seen in the phase diagram under biaxial strain. 

The RP structures bear both additional degrees of freedom (layering) and structural complexity compared to ABO$_3$ perovskites. This study provides an example of theoretical determination of the RP oxide structure, analytical understanding of stability, and phase transitions under different boundary conditions and shows that even in the absence of additional complications (such as Jahn-Teller distortions, magnetic transitions, etc.) the phase diagrams can host multiple phases. The effect of these structural phases on electronic properties such as the bandgap and the effective mass will enable tuning of material properties and broadening of its applicability in opto-electronic devices.

\begin{acknowledgments}
This work was supported primarily by the National Science Foundation through the University of Minnesota MRSEC under Award Number DMR-2011401. The authors acknowledge the Minnesota Supercomputing Institute (MSI) at the University of Minnesota for providing computational resources that contributed to the research results reported within this paper.
\end{acknowledgments}


\begin{thebibliography}{68}%
\makeatletter
\providecommand \@ifxundefined [1]{%
 \@ifx{#1\undefined}
}%
\providecommand \@ifnum [1]{%
 \ifnum #1\expandafter \@firstoftwo
 \else \expandafter \@secondoftwo
 \fi
}%
\providecommand \@ifx [1]{%
 \ifx #1\expandafter \@firstoftwo
 \else \expandafter \@secondoftwo
 \fi
}%
\providecommand \natexlab [1]{#1}%
\providecommand \enquote  [1]{``#1''}%
\providecommand \bibnamefont  [1]{#1}%
\providecommand \bibfnamefont [1]{#1}%
\providecommand \citenamefont [1]{#1}%
\providecommand \href@noop [0]{\@secondoftwo}%
\providecommand \href [0]{\begingroup \@sanitize@url \@href}%
\providecommand \@href[1]{\@@startlink{#1}\@@href}%
\providecommand \@@href[1]{\endgroup#1\@@endlink}%
\providecommand \@sanitize@url [0]{\catcode `\\12\catcode `\$12\catcode
  `\&12\catcode `\#12\catcode `\^12\catcode `\_12\catcode `\%12\relax}%
\providecommand \@@startlink[1]{}%
\providecommand \@@endlink[0]{}%
\providecommand \url  [0]{\begingroup\@sanitize@url \@url }%
\providecommand \@url [1]{\endgroup\@href {#1}{\urlprefix }}%
\providecommand \urlprefix  [0]{URL }%
\providecommand \Eprint [0]{\href }%
\providecommand \doibase [0]{http://dx.doi.org/}%
\providecommand \selectlanguage [0]{\@gobble}%
\providecommand \bibinfo  [0]{\@secondoftwo}%
\providecommand \bibfield  [0]{\@secondoftwo}%
\providecommand \translation [1]{[#1]}%
\providecommand \BibitemOpen [0]{}%
\providecommand \bibitemStop [0]{}%
\providecommand \bibitemNoStop [0]{.\EOS\space}%
\providecommand \EOS [0]{\spacefactor3000\relax}%
\providecommand \BibitemShut  [1]{\csname bibitem#1\endcsname}%
\let\auto@bib@innerbib\@empty
\bibitem [{\citenamefont {Lufaso}\ and\ \citenamefont
  {Woodward}(2001)}]{Lufaso2001}%
  \BibitemOpen
  \bibfield  {author} {\bibinfo {author} {\bibfnamefont {M.~W.}\ \bibnamefont
  {Lufaso}}\ and\ \bibinfo {author} {\bibfnamefont {P.~M.}\ \bibnamefont
  {Woodward}},\ }\href@noop {} {\bibfield  {journal} {\bibinfo  {journal} {Acta
  Crystallogr. B}\ }\textbf {\bibinfo {volume} {57}},\ \bibinfo {pages} {725}
  (\bibinfo {year} {2001})}\BibitemShut {NoStop}%
\bibitem [{\citenamefont {Tilley}(2008)}]{Tilley2008}%
  \BibitemOpen
  \bibfield  {author} {\bibinfo {author} {\bibfnamefont {R.}~\bibnamefont
  {Tilley}},\ }\href@noop {} {\emph {\bibinfo {title} {Defects in Solids}}},\
  Special Topics in Inorganic Chemistry\ (\bibinfo  {publisher} {Wiley},\
  \bibinfo {year} {2008})\BibitemShut {NoStop}%
\bibitem [{\citenamefont {Tilley}(1977)}]{Tilley1977}%
  \BibitemOpen
  \bibfield  {author} {\bibinfo {author} {\bibfnamefont {R.~J.}\ \bibnamefont
  {Tilley}},\ }\href {\doibase 10.1016/0022-4596(77)90128-1} {\bibfield
  {journal} {\bibinfo  {journal} {J. Solid State Chem.}\ }\textbf {\bibinfo
  {volume} {21}},\ \bibinfo {pages} {293} (\bibinfo {year} {1977})}\BibitemShut
  {NoStop}%
\bibitem [{\citenamefont {Ruddlesden}\ and\ \citenamefont
  {Popper}(1957)}]{Ruddlesden1957}%
  \BibitemOpen
  \bibfield  {author} {\bibinfo {author} {\bibfnamefont {S.~N.}\ \bibnamefont
  {Ruddlesden}}\ and\ \bibinfo {author} {\bibfnamefont {P.}~\bibnamefont
  {Popper}},\ }\href {\doibase 10.1107/S0365110X57001929} {\bibfield  {journal}
  {\bibinfo  {journal} {Acta Crystallogr.}\ }\textbf {\bibinfo {volume} {10}},\
  \bibinfo {pages} {538} (\bibinfo {year} {1957})}\BibitemShut {NoStop}%
\bibitem [{\citenamefont {Ruddlesden}\ and\ \citenamefont
  {Popper}(1958)}]{Ruddlesden1958}%
  \BibitemOpen
  \bibfield  {author} {\bibinfo {author} {\bibfnamefont {S.~N.}\ \bibnamefont
  {Ruddlesden}}\ and\ \bibinfo {author} {\bibfnamefont {P.}~\bibnamefont
  {Popper}},\ }\href {\doibase 10.1107/S0365110X58000128} {\bibfield  {journal}
  {\bibinfo  {journal} {Acta Crystallogr.}\ }\textbf {\bibinfo {volume} {11}},\
  \bibinfo {pages} {54} (\bibinfo {year} {1958})}\BibitemShut {NoStop}%
\bibitem [{\citenamefont {Moon}\ \emph {et~al.}(2008)\citenamefont {Moon},
  \citenamefont {Jin}, \citenamefont {Kim}, \citenamefont {Choi}, \citenamefont
  {Lee}, \citenamefont {Yu}, \citenamefont {Cao}, \citenamefont {Sumi},
  \citenamefont {Funakubo}, \citenamefont {Bernhard},\ and\ \citenamefont
  {Noh}}]{Moon2008}%
  \BibitemOpen
  \bibfield  {author} {\bibinfo {author} {\bibfnamefont {S.~J.}\ \bibnamefont
  {Moon}}, \bibinfo {author} {\bibfnamefont {H.}~\bibnamefont {Jin}}, \bibinfo
  {author} {\bibfnamefont {K.~W.}\ \bibnamefont {Kim}}, \bibinfo {author}
  {\bibfnamefont {W.~S.}\ \bibnamefont {Choi}}, \bibinfo {author}
  {\bibfnamefont {Y.~S.}\ \bibnamefont {Lee}}, \bibinfo {author} {\bibfnamefont
  {J.}~\bibnamefont {Yu}}, \bibinfo {author} {\bibfnamefont {G.}~\bibnamefont
  {Cao}}, \bibinfo {author} {\bibfnamefont {A.}~\bibnamefont {Sumi}}, \bibinfo
  {author} {\bibfnamefont {H.}~\bibnamefont {Funakubo}}, \bibinfo {author}
  {\bibfnamefont {C.}~\bibnamefont {Bernhard}}, \ and\ \bibinfo {author}
  {\bibfnamefont {T.~W.}\ \bibnamefont {Noh}},\ }\href@noop {} {\bibfield
  {journal} {\bibinfo  {journal} {Phys. Rev. Lett.}\ }\textbf {\bibinfo
  {volume} {101}},\ \bibinfo {pages} {226402} (\bibinfo {year}
  {2008})}\BibitemShut {NoStop}%
\bibitem [{\citenamefont {Mulder}\ \emph {et~al.}(2013)\citenamefont {Mulder},
  \citenamefont {Benedek}, \citenamefont {Rondinelli},\ and\ \citenamefont
  {Fennie}}]{Mulder2013}%
  \BibitemOpen
  \bibfield  {author} {\bibinfo {author} {\bibfnamefont {A.~T.}\ \bibnamefont
  {Mulder}}, \bibinfo {author} {\bibfnamefont {N.~A.}\ \bibnamefont {Benedek}},
  \bibinfo {author} {\bibfnamefont {J.~M.}\ \bibnamefont {Rondinelli}}, \ and\
  \bibinfo {author} {\bibfnamefont {C.~J.}\ \bibnamefont {Fennie}},\
  }\href@noop {} {\bibfield  {journal} {\bibinfo  {journal} {Adv. Funct.
  Mater.}\ }\textbf {\bibinfo {volume} {23}},\ \bibinfo {pages} {4810}
  (\bibinfo {year} {2013})}\BibitemShut {NoStop}%
\bibitem [{\citenamefont {Birol}\ \emph {et~al.}(2011)\citenamefont {Birol},
  \citenamefont {Benedek},\ and\ \citenamefont {Fennie}}]{Birol2011}%
  \BibitemOpen
  \bibfield  {author} {\bibinfo {author} {\bibfnamefont {T.}~\bibnamefont
  {Birol}}, \bibinfo {author} {\bibfnamefont {N.~A.}\ \bibnamefont {Benedek}},
  \ and\ \bibinfo {author} {\bibfnamefont {C.~J.}\ \bibnamefont {Fennie}},\
  }\href@noop {} {\bibfield  {journal} {\bibinfo  {journal} {Phys. Rev. Lett.}\
  }\textbf {\bibinfo {volume} {107}},\ \bibinfo {pages} {257602} (\bibinfo
  {year} {2011})}\BibitemShut {NoStop}%
\bibitem [{\citenamefont {Woodward}(1997{\natexlab{a}})}]{Woodward1997a}%
  \BibitemOpen
  \bibfield  {author} {\bibinfo {author} {\bibfnamefont {P.~M.}\ \bibnamefont
  {Woodward}},\ }\href@noop {} {\bibfield  {journal} {\bibinfo  {journal} {Acta
  Crystallogr. B}\ }\textbf {\bibinfo {volume} {53}},\ \bibinfo {pages} {32}
  (\bibinfo {year} {1997}{\natexlab{a}})}\BibitemShut {NoStop}%
\bibitem [{\citenamefont {Balachandran}\ \emph {et~al.}(2014)\citenamefont
  {Balachandran}, \citenamefont {Puggioni},\ and\ \citenamefont
  {Rondinelli}}]{Balachandran2013}%
  \BibitemOpen
  \bibfield  {author} {\bibinfo {author} {\bibfnamefont {P.~V.}\ \bibnamefont
  {Balachandran}}, \bibinfo {author} {\bibfnamefont {D.}~\bibnamefont
  {Puggioni}}, \ and\ \bibinfo {author} {\bibfnamefont {J.~M.}\ \bibnamefont
  {Rondinelli}},\ }\href@noop {} {\bibfield  {journal} {\bibinfo  {journal}
  {Inorg. Chem.}\ }\textbf {\bibinfo {volume} {53}},\ \bibinfo {pages} {336}
  (\bibinfo {year} {2014})}\BibitemShut {NoStop}%
\bibitem [{\citenamefont {Li}\ and\ \citenamefont {Birol}(2020)}]{Li2020}%
  \BibitemOpen
  \bibfield  {author} {\bibinfo {author} {\bibfnamefont {S.}~\bibnamefont
  {Li}}\ and\ \bibinfo {author} {\bibfnamefont {T.}~\bibnamefont {Birol}},\
  }\href@noop {} {\bibfield  {journal} {\bibinfo  {journal} {npj Comput.
  Mater.}\ }\textbf {\bibinfo {volume} {6}} (\bibinfo {year}
  {2020})}\BibitemShut {NoStop}%
\bibitem [{\citenamefont {Lee}\ \emph {et~al.}(2013)\citenamefont {Lee},
  \citenamefont {Podraza}, \citenamefont {Zhu}, \citenamefont {Berger},
  \citenamefont {Shen}, \citenamefont {Sestak}, \citenamefont {Collins},
  \citenamefont {Kourkoutis}, \citenamefont {Mundy}, \citenamefont {Wang},
  \citenamefont {Mao}, \citenamefont {Xi}, \citenamefont {Brillson},
  \citenamefont {Neaton}, \citenamefont {Muller},\ and\ \citenamefont
  {Schlom}}]{Lee2013}%
  \BibitemOpen
  \bibfield  {author} {\bibinfo {author} {\bibfnamefont {C.-H.}\ \bibnamefont
  {Lee}}, \bibinfo {author} {\bibfnamefont {N.~J.}\ \bibnamefont {Podraza}},
  \bibinfo {author} {\bibfnamefont {Y.}~\bibnamefont {Zhu}}, \bibinfo {author}
  {\bibfnamefont {R.~F.}\ \bibnamefont {Berger}}, \bibinfo {author}
  {\bibfnamefont {S.}~\bibnamefont {Shen}}, \bibinfo {author} {\bibfnamefont
  {M.}~\bibnamefont {Sestak}}, \bibinfo {author} {\bibfnamefont {R.~W.}\
  \bibnamefont {Collins}}, \bibinfo {author} {\bibfnamefont {L.~F.}\
  \bibnamefont {Kourkoutis}}, \bibinfo {author} {\bibfnamefont {J.~A.}\
  \bibnamefont {Mundy}}, \bibinfo {author} {\bibfnamefont {H.}~\bibnamefont
  {Wang}}, \bibinfo {author} {\bibfnamefont {Q.}~\bibnamefont {Mao}}, \bibinfo
  {author} {\bibfnamefont {X.}~\bibnamefont {Xi}}, \bibinfo {author}
  {\bibfnamefont {L.~J.}\ \bibnamefont {Brillson}}, \bibinfo {author}
  {\bibfnamefont {J.~B.}\ \bibnamefont {Neaton}}, \bibinfo {author}
  {\bibfnamefont {D.~A.}\ \bibnamefont {Muller}}, \ and\ \bibinfo {author}
  {\bibfnamefont {D.~G.}\ \bibnamefont {Schlom}},\ }\href@noop {} {\bibfield
  {journal} {\bibinfo  {journal} {Appl. Phys. Lett.}\ }\textbf {\bibinfo
  {volume} {102}},\ \bibinfo {pages} {122901} (\bibinfo {year}
  {2013})}\BibitemShut {NoStop}%
\bibitem [{\citenamefont {Ramkumar}\ and\ \citenamefont
  {Nowadnick}(2021)}]{Ramkumar2021}%
  \BibitemOpen
  \bibfield  {author} {\bibinfo {author} {\bibfnamefont {S.~P.}\ \bibnamefont
  {Ramkumar}}\ and\ \bibinfo {author} {\bibfnamefont {E.~A.}\ \bibnamefont
  {Nowadnick}},\ }\href@noop {} {\bibfield  {journal} {\bibinfo  {journal}
  {Phys. Rev. B}\ }\textbf {\bibinfo {volume} {104}},\ \bibinfo {pages}
  {144105} (\bibinfo {year} {2021})}\BibitemShut {NoStop}%
\bibitem [{\citenamefont {Lu}\ and\ \citenamefont {Rondinelli}(2016)}]{Lu2016}%
  \BibitemOpen
  \bibfield  {author} {\bibinfo {author} {\bibfnamefont {X.-Z.}\ \bibnamefont
  {Lu}}\ and\ \bibinfo {author} {\bibfnamefont {J.~M.}\ \bibnamefont
  {Rondinelli}},\ }\href@noop {} {\bibfield  {journal} {\bibinfo  {journal}
  {Nat. Mater.}\ }\textbf {\bibinfo {volume} {15}},\ \bibinfo {pages} {951}
  (\bibinfo {year} {2016})}\BibitemShut {NoStop}%
\bibitem [{\citenamefont {Glerup}\ \emph {et~al.}(2005)\citenamefont {Glerup},
  \citenamefont {Knight},\ and\ \citenamefont {Poulsen}}]{Glerup2005}%
  \BibitemOpen
  \bibfield  {author} {\bibinfo {author} {\bibfnamefont {M.}~\bibnamefont
  {Glerup}}, \bibinfo {author} {\bibfnamefont {K.~S.}\ \bibnamefont {Knight}},
  \ and\ \bibinfo {author} {\bibfnamefont {F.~W.}\ \bibnamefont {Poulsen}},\
  }\href@noop {} {\bibfield  {journal} {\bibinfo  {journal} {Mater. Res.
  Bull.}\ }\textbf {\bibinfo {volume} {40}},\ \bibinfo {pages} {507} (\bibinfo
  {year} {2005})}\BibitemShut {NoStop}%
\bibitem [{\citenamefont {Zhang}\ \emph {et~al.}(2017)\citenamefont {Zhang},
  \citenamefont {Wang}, \citenamefont {Sahoo}, \citenamefont {Shimada},\ and\
  \citenamefont {Kitamura}}]{Zhang2017}%
  \BibitemOpen
  \bibfield  {author} {\bibinfo {author} {\bibfnamefont {Y.}~\bibnamefont
  {Zhang}}, \bibinfo {author} {\bibfnamefont {J.}~\bibnamefont {Wang}},
  \bibinfo {author} {\bibfnamefont {M.~P.~K.}\ \bibnamefont {Sahoo}}, \bibinfo
  {author} {\bibfnamefont {T.}~\bibnamefont {Shimada}}, \ and\ \bibinfo
  {author} {\bibfnamefont {T.}~\bibnamefont {Kitamura}},\ }\href@noop {}
  {\bibfield  {journal} {\bibinfo  {journal} {Phys. Chem. Chem. Phys.}\
  }\textbf {\bibinfo {volume} {19}},\ \bibinfo {pages} {26047} (\bibinfo {year}
  {2017})}\BibitemShut {NoStop}%
\bibitem [{\citenamefont {Wang}\ \emph {et~al.}(2018)\citenamefont {Wang},
  \citenamefont {Prakash}, \citenamefont {Dong}, \citenamefont {Truttmann},
  \citenamefont {Bucsek}, \citenamefont {James}, \citenamefont {Fong},
  \citenamefont {Kim}, \citenamefont {Ryan}, \citenamefont {Zhou},
  \citenamefont {Birol},\ and\ \citenamefont {Jalan}}]{Wang2018}%
  \BibitemOpen
  \bibfield  {author} {\bibinfo {author} {\bibfnamefont {T.}~\bibnamefont
  {Wang}}, \bibinfo {author} {\bibfnamefont {A.}~\bibnamefont {Prakash}},
  \bibinfo {author} {\bibfnamefont {Y.}~\bibnamefont {Dong}}, \bibinfo {author}
  {\bibfnamefont {T.}~\bibnamefont {Truttmann}}, \bibinfo {author}
  {\bibfnamefont {A.}~\bibnamefont {Bucsek}}, \bibinfo {author} {\bibfnamefont
  {R.}~\bibnamefont {James}}, \bibinfo {author} {\bibfnamefont {D.~D.}\
  \bibnamefont {Fong}}, \bibinfo {author} {\bibfnamefont {J.-W.}\ \bibnamefont
  {Kim}}, \bibinfo {author} {\bibfnamefont {P.~J.}\ \bibnamefont {Ryan}},
  \bibinfo {author} {\bibfnamefont {H.}~\bibnamefont {Zhou}}, \bibinfo {author}
  {\bibfnamefont {T.}~\bibnamefont {Birol}}, \ and\ \bibinfo {author}
  {\bibfnamefont {B.}~\bibnamefont {Jalan}},\ }\href@noop {} {\bibfield
  {journal} {\bibinfo  {journal} {ACS Appl. Mater. Interfaces}\ }\textbf
  {\bibinfo {volume} {10}},\ \bibinfo {pages} {43802} (\bibinfo {year}
  {2018})}\BibitemShut {NoStop}%
\bibitem [{\citenamefont {Prakash}\ \emph {et~al.}(2021)\citenamefont
  {Prakash}, \citenamefont {Wang}, \citenamefont {Bucsek}, \citenamefont
  {Truttmann}, \citenamefont {Fali}, \citenamefont {Cotrufo}, \citenamefont
  {Yun}, \citenamefont {Kim}, \citenamefont {Ryan}, \citenamefont {Mkhoyan},
  \citenamefont {Alù}, \citenamefont {Abate}, \citenamefont {James},\ and\
  \citenamefont {Jalan}}]{Prakash2021}%
  \BibitemOpen
  \bibfield  {author} {\bibinfo {author} {\bibfnamefont {A.}~\bibnamefont
  {Prakash}}, \bibinfo {author} {\bibfnamefont {T.}~\bibnamefont {Wang}},
  \bibinfo {author} {\bibfnamefont {A.}~\bibnamefont {Bucsek}}, \bibinfo
  {author} {\bibfnamefont {T.~K.}\ \bibnamefont {Truttmann}}, \bibinfo {author}
  {\bibfnamefont {A.}~\bibnamefont {Fali}}, \bibinfo {author} {\bibfnamefont
  {M.}~\bibnamefont {Cotrufo}}, \bibinfo {author} {\bibfnamefont
  {H.}~\bibnamefont {Yun}}, \bibinfo {author} {\bibfnamefont {J.-W.}\
  \bibnamefont {Kim}}, \bibinfo {author} {\bibfnamefont {P.~J.}\ \bibnamefont
  {Ryan}}, \bibinfo {author} {\bibfnamefont {K.~A.}\ \bibnamefont {Mkhoyan}},
  \bibinfo {author} {\bibfnamefont {A.}~\bibnamefont {Alù}}, \bibinfo {author}
  {\bibfnamefont {Y.}~\bibnamefont {Abate}}, \bibinfo {author} {\bibfnamefont
  {R.~D.}\ \bibnamefont {James}}, \ and\ \bibinfo {author} {\bibfnamefont
  {B.}~\bibnamefont {Jalan}},\ }\href@noop {} {\bibfield  {journal} {\bibinfo
  {journal} {Nano Letters}\ }\textbf {\bibinfo {volume} {21}},\ \bibinfo
  {pages} {1246} (\bibinfo {year} {2021})}\BibitemShut {NoStop}%
\bibitem [{\citenamefont {Kim}\ \emph {et~al.}(2022)\citenamefont {Kim},
  \citenamefont {Yun}, \citenamefont {Seo}, \citenamefont {Kim}, \citenamefont
  {Kim}, \citenamefont {Mkhoyan}, \citenamefont {Kim},\ and\ \citenamefont
  {Char}}]{Kim2022}%
  \BibitemOpen
  \bibfield  {author} {\bibinfo {author} {\bibfnamefont {J.}~\bibnamefont
  {Kim}}, \bibinfo {author} {\bibfnamefont {H.}~\bibnamefont {Yun}}, \bibinfo
  {author} {\bibfnamefont {J.}~\bibnamefont {Seo}}, \bibinfo {author}
  {\bibfnamefont {J.~H.}\ \bibnamefont {Kim}}, \bibinfo {author} {\bibfnamefont
  {J.~H.}\ \bibnamefont {Kim}}, \bibinfo {author} {\bibfnamefont {K.~A.}\
  \bibnamefont {Mkhoyan}}, \bibinfo {author} {\bibfnamefont {B.}~\bibnamefont
  {Kim}}, \ and\ \bibinfo {author} {\bibfnamefont {K.}~\bibnamefont {Char}},\
  }\href@noop {} {\bibfield  {journal} {\bibinfo  {journal} {ACS Appl. Electr.
  Mater.}\ }\textbf {\bibinfo {volume} {4}},\ \bibinfo {pages} {3623} (\bibinfo
  {year} {2022})}\BibitemShut {NoStop}%
\bibitem [{\citenamefont {Green}\ \emph {et~al.}(1996)\citenamefont {Green},
  \citenamefont {Prassides}, \citenamefont {Day},\ and\ \citenamefont
  {Stalick}}]{Green1996}%
  \BibitemOpen
  \bibfield  {author} {\bibinfo {author} {\bibfnamefont {M.~A.}\ \bibnamefont
  {Green}}, \bibinfo {author} {\bibfnamefont {K.}~\bibnamefont {Prassides}},
  \bibinfo {author} {\bibfnamefont {P.}~\bibnamefont {Day}}, \ and\ \bibinfo
  {author} {\bibfnamefont {J.~K.}\ \bibnamefont {Stalick}},\ }\href@noop {}
  {\bibfield  {journal} {\bibinfo  {journal} {J. Chem. Soc., Faraday Trans.}\
  }\textbf {\bibinfo {volume} {92}},\ \bibinfo {pages} {2155} (\bibinfo {year}
  {1996})}\BibitemShut {NoStop}%
\bibitem [{\citenamefont {Green}\ \emph {et~al.}(2000)\citenamefont {Green},
  \citenamefont {Prassides}, \citenamefont {Day},\ and\ \citenamefont
  {Neumann}}]{Green2000}%
  \BibitemOpen
  \bibfield  {author} {\bibinfo {author} {\bibfnamefont {M.}~\bibnamefont
  {Green}}, \bibinfo {author} {\bibfnamefont {K.}~\bibnamefont {Prassides}},
  \bibinfo {author} {\bibfnamefont {P.}~\bibnamefont {Day}}, \ and\ \bibinfo
  {author} {\bibfnamefont {D.}~\bibnamefont {Neumann}},\ }\href@noop {}
  {\bibfield  {journal} {\bibinfo  {journal} {Int. J. Inorg. Mater.}\ }\textbf
  {\bibinfo {volume} {2}},\ \bibinfo {pages} {35} (\bibinfo {year}
  {2000})}\BibitemShut {NoStop}%
\bibitem [{\citenamefont {Ueda}\ \emph {et~al.}(2006)\citenamefont {Ueda},
  \citenamefont {Yamashita}, \citenamefont {Nakayashiki}, \citenamefont {Goto},
  \citenamefont {Maeda}, \citenamefont {Furui}, \citenamefont {Ozaki},
  \citenamefont {Nakachi}, \citenamefont {Nakamura}, \citenamefont {Fujisawa},\
  and\ \citenamefont {Miyazaki}}]{Ueda2006}%
  \BibitemOpen
  \bibfield  {author} {\bibinfo {author} {\bibfnamefont {K.}~\bibnamefont
  {Ueda}}, \bibinfo {author} {\bibfnamefont {T.}~\bibnamefont {Yamashita}},
  \bibinfo {author} {\bibfnamefont {K.}~\bibnamefont {Nakayashiki}}, \bibinfo
  {author} {\bibfnamefont {K.}~\bibnamefont {Goto}}, \bibinfo {author}
  {\bibfnamefont {T.}~\bibnamefont {Maeda}}, \bibinfo {author} {\bibfnamefont
  {K.}~\bibnamefont {Furui}}, \bibinfo {author} {\bibfnamefont
  {K.}~\bibnamefont {Ozaki}}, \bibinfo {author} {\bibfnamefont
  {Y.}~\bibnamefont {Nakachi}}, \bibinfo {author} {\bibfnamefont
  {S.}~\bibnamefont {Nakamura}}, \bibinfo {author} {\bibfnamefont
  {M.}~\bibnamefont {Fujisawa}}, \ and\ \bibinfo {author} {\bibfnamefont
  {T.}~\bibnamefont {Miyazaki}},\ }\href@noop {} {\bibfield  {journal}
  {\bibinfo  {journal} {Jpn. J. Appl. Phys.}\ }\textbf {\bibinfo {volume}
  {45}},\ \bibinfo {pages} {6981} (\bibinfo {year} {2006})}\BibitemShut
  {NoStop}%
\bibitem [{\citenamefont {Kamimura}\ \emph {et~al.}(2014)\citenamefont
  {Kamimura}, \citenamefont {Xu}, \citenamefont {Yamada}, \citenamefont
  {Terasaki},\ and\ \citenamefont {Fujihala}}]{Kamimura2014}%
  \BibitemOpen
  \bibfield  {author} {\bibinfo {author} {\bibfnamefont {S.}~\bibnamefont
  {Kamimura}}, \bibinfo {author} {\bibfnamefont {C.-N.}\ \bibnamefont {Xu}},
  \bibinfo {author} {\bibfnamefont {H.}~\bibnamefont {Yamada}}, \bibinfo
  {author} {\bibfnamefont {N.}~\bibnamefont {Terasaki}}, \ and\ \bibinfo
  {author} {\bibfnamefont {M.}~\bibnamefont {Fujihala}},\ }\href@noop {}
  {\bibfield  {journal} {\bibinfo  {journal} {Jpn. J. Appl. Phys.}\ }\textbf
  {\bibinfo {volume} {53}},\ \bibinfo {pages} {092403} (\bibinfo {year}
  {2014})}\BibitemShut {NoStop}%
\bibitem [{\citenamefont {Kamimura}\ \emph {et~al.}(2012)\citenamefont
  {Kamimura}, \citenamefont {Yamada},\ and\ \citenamefont {Xu}}]{Kamimura2012}%
  \BibitemOpen
  \bibfield  {author} {\bibinfo {author} {\bibfnamefont {S.}~\bibnamefont
  {Kamimura}}, \bibinfo {author} {\bibfnamefont {H.}~\bibnamefont {Yamada}}, \
  and\ \bibinfo {author} {\bibfnamefont {C.-N.}\ \bibnamefont {Xu}},\
  }\href@noop {} {\bibfield  {journal} {\bibinfo  {journal} {Appl. Phys.
  Lett.}\ }\textbf {\bibinfo {volume} {101}},\ \bibinfo {pages} {091113}
  (\bibinfo {year} {2012})}\BibitemShut {NoStop}%
\bibitem [{\citenamefont {Wang}\ \emph {et~al.}(2020)\citenamefont {Wang},
  \citenamefont {Zheng}, \citenamefont {Zhang}, \citenamefont {Liu},
  \citenamefont {Deng}, \citenamefont {Xu}, \citenamefont {Zhou},\ and\
  \citenamefont {He}}]{Wang2020}%
  \BibitemOpen
  \bibfield  {author} {\bibinfo {author} {\bibfnamefont {C.}~\bibnamefont
  {Wang}}, \bibinfo {author} {\bibfnamefont {Z.}~\bibnamefont {Zheng}},
  \bibinfo {author} {\bibfnamefont {Y.}~\bibnamefont {Zhang}}, \bibinfo
  {author} {\bibfnamefont {Q.}~\bibnamefont {Liu}}, \bibinfo {author}
  {\bibfnamefont {M.}~\bibnamefont {Deng}}, \bibinfo {author} {\bibfnamefont
  {X.}~\bibnamefont {Xu}}, \bibinfo {author} {\bibfnamefont {Z.}~\bibnamefont
  {Zhou}}, \ and\ \bibinfo {author} {\bibfnamefont {H.}~\bibnamefont {He}},\
  }\href@noop {} {\bibfield  {journal} {\bibinfo  {journal} {Opt. Express}\
  }\textbf {\bibinfo {volume} {28}},\ \bibinfo {pages} {4249} (\bibinfo {year}
  {2020})}\BibitemShut {NoStop}%
\bibitem [{\citenamefont {Kumar}\ and\ \citenamefont
  {Upadhyay}(2020)}]{Kumar2020}%
  \BibitemOpen
  \bibfield  {author} {\bibinfo {author} {\bibfnamefont {U.}~\bibnamefont
  {Kumar}}\ and\ \bibinfo {author} {\bibfnamefont {S.}~\bibnamefont
  {Upadhyay}},\ }\href@noop {} {\bibfield  {journal} {\bibinfo  {journal} {J.
  Mater. Sci.: Mater. Electron.}\ }\textbf {\bibinfo {volume} {31}} (\bibinfo
  {year} {2020})}\BibitemShut {NoStop}%
\bibitem [{\citenamefont {Nirala}\ \emph {et~al.}(2020)\citenamefont {Nirala},
  \citenamefont {Yadav},\ and\ \citenamefont {Upadhyay}}]{Nirala2020}%
  \BibitemOpen
  \bibfield  {author} {\bibinfo {author} {\bibfnamefont {G.}~\bibnamefont
  {Nirala}}, \bibinfo {author} {\bibfnamefont {D.}~\bibnamefont {Yadav}}, \
  and\ \bibinfo {author} {\bibfnamefont {S.}~\bibnamefont {Upadhyay}},\
  }\href@noop {} {\bibfield  {journal} {\bibinfo  {journal} {J. Adv. Ceram.}\
  }\textbf {\bibinfo {volume} {9}} (\bibinfo {year} {2020})}\BibitemShut
  {NoStop}%
\bibitem [{\citenamefont {Kumar}\ and\ \citenamefont
  {Upadhyay}(2018)}]{Kumar2018}%
  \BibitemOpen
  \bibfield  {author} {\bibinfo {author} {\bibfnamefont {U.}~\bibnamefont
  {Kumar}}\ and\ \bibinfo {author} {\bibfnamefont {S.}~\bibnamefont
  {Upadhyay}},\ }\href@noop {} {\bibfield  {journal} {\bibinfo  {journal}
  {Mater. Lett.}\ }\textbf {\bibinfo {volume} {227}},\ \bibinfo {pages} {100}
  (\bibinfo {year} {2018})}\BibitemShut {NoStop}%
\bibitem [{\citenamefont {Fu}\ \emph {et~al.}(2002)\citenamefont {Fu},
  \citenamefont {Visser},\ and\ \citenamefont {IJdo}}]{Fu2002}%
  \BibitemOpen
  \bibfield  {author} {\bibinfo {author} {\bibfnamefont {W.}~\bibnamefont
  {Fu}}, \bibinfo {author} {\bibfnamefont {D.}~\bibnamefont {Visser}}, \ and\
  \bibinfo {author} {\bibfnamefont {D.}~\bibnamefont {IJdo}},\ }\href@noop {}
  {\bibfield  {journal} {\bibinfo  {journal} {J. Solid State Chem.}\ }\textbf
  {\bibinfo {volume} {169}},\ \bibinfo {pages} {208} (\bibinfo {year}
  {2002})}\BibitemShut {NoStop}%
\bibitem [{\citenamefont {Fu}\ \emph {et~al.}(2004)\citenamefont {Fu},
  \citenamefont {Visser}, \citenamefont {Knight},\ and\ \citenamefont
  {IJdo}}]{Fu2004}%
  \BibitemOpen
  \bibfield  {author} {\bibinfo {author} {\bibfnamefont {W.}~\bibnamefont
  {Fu}}, \bibinfo {author} {\bibfnamefont {D.}~\bibnamefont {Visser}}, \bibinfo
  {author} {\bibfnamefont {K.}~\bibnamefont {Knight}}, \ and\ \bibinfo {author}
  {\bibfnamefont {D.}~\bibnamefont {IJdo}},\ }\href@noop {} {\bibfield
  {journal} {\bibinfo  {journal} {J. Solid State Chem.}\ }\textbf {\bibinfo
  {volume} {177}},\ \bibinfo {pages} {4081} (\bibinfo {year}
  {2004})}\BibitemShut {NoStop}%
\bibitem [{\citenamefont {Balachandran}\ and\ \citenamefont
  {Rondinelli}(2013)}]{Balachandran2013_2}%
  \BibitemOpen
  \bibfield  {author} {\bibinfo {author} {\bibfnamefont {P.~V.}\ \bibnamefont
  {Balachandran}}\ and\ \bibinfo {author} {\bibfnamefont {J.~M.}\ \bibnamefont
  {Rondinelli}},\ }\href@noop {} {\bibfield  {journal} {\bibinfo  {journal}
  {Phys. Rev. B}\ }\textbf {\bibinfo {volume} {88}},\ \bibinfo {pages} {054101}
  (\bibinfo {year} {2013})}\BibitemShut {NoStop}%
\bibitem [{\citenamefont {Stokes}\ \emph {et~al.}(2002)\citenamefont {Stokes},
  \citenamefont {Kisi}, \citenamefont {Hatch},\ and\ \citenamefont
  {Howard}}]{Stokes2002}%
  \BibitemOpen
  \bibfield  {author} {\bibinfo {author} {\bibfnamefont {H.~T.}\ \bibnamefont
  {Stokes}}, \bibinfo {author} {\bibfnamefont {E.~H.}\ \bibnamefont {Kisi}},
  \bibinfo {author} {\bibfnamefont {D.~M.}\ \bibnamefont {Hatch}}, \ and\
  \bibinfo {author} {\bibfnamefont {C.~J.}\ \bibnamefont {Howard}},\
  }\href@noop {} {\bibfield  {journal} {\bibinfo  {journal} {Acta Crystallogr.
  B}\ }\textbf {\bibinfo {volume} {58}},\ \bibinfo {pages} {934} (\bibinfo
  {year} {2002})}\BibitemShut {NoStop}%
\bibitem [{\citenamefont {Schlom}\ \emph {et~al.}(2007)\citenamefont {Schlom},
  \citenamefont {Chen}, \citenamefont {Eom}, \citenamefont {Rabe},
  \citenamefont {Streiffer},\ and\ \citenamefont {Triscone}}]{Schlom2007}%
  \BibitemOpen
  \bibfield  {author} {\bibinfo {author} {\bibfnamefont {D.~G.}\ \bibnamefont
  {Schlom}}, \bibinfo {author} {\bibfnamefont {L.-Q.}\ \bibnamefont {Chen}},
  \bibinfo {author} {\bibfnamefont {C.-B.}\ \bibnamefont {Eom}}, \bibinfo
  {author} {\bibfnamefont {K.~M.}\ \bibnamefont {Rabe}}, \bibinfo {author}
  {\bibfnamefont {S.~K.}\ \bibnamefont {Streiffer}}, \ and\ \bibinfo {author}
  {\bibfnamefont {J.-M.}\ \bibnamefont {Triscone}},\ }\href {\doibase
  10.1146/annurev.matsci.37.061206.113016} {\bibfield  {journal} {\bibinfo
  {journal} {Annu. Rev. Mater. Res.}\ }\textbf {\bibinfo {volume} {37}},\
  \bibinfo {pages} {589} (\bibinfo {year} {2007})}\BibitemShut {NoStop}%
\bibitem [{\citenamefont {Schlom}\ \emph {et~al.}(2014)\citenamefont {Schlom},
  \citenamefont {Chen}, \citenamefont {Fennie}, \citenamefont {Gopalan},
  \citenamefont {Muller}, \citenamefont {Pan}, \citenamefont {Ramesh},\ and\
  \citenamefont {Uecker}}]{Schlom2014}%
  \BibitemOpen
  \bibfield  {author} {\bibinfo {author} {\bibfnamefont {D.~G.}\ \bibnamefont
  {Schlom}}, \bibinfo {author} {\bibfnamefont {L.~Q.}\ \bibnamefont {Chen}},
  \bibinfo {author} {\bibfnamefont {C.~J.}\ \bibnamefont {Fennie}}, \bibinfo
  {author} {\bibfnamefont {V.}~\bibnamefont {Gopalan}}, \bibinfo {author}
  {\bibfnamefont {D.~A.}\ \bibnamefont {Muller}}, \bibinfo {author}
  {\bibfnamefont {X.}~\bibnamefont {Pan}}, \bibinfo {author} {\bibfnamefont
  {R.}~\bibnamefont {Ramesh}}, \ and\ \bibinfo {author} {\bibfnamefont
  {R.}~\bibnamefont {Uecker}},\ }\href {\doibase 10.1557/mrs.2014.1} {\bibfield
   {journal} {\bibinfo  {journal} {MRS Bulletin}\ }\textbf {\bibinfo {volume}
  {39}},\ \bibinfo {pages} {118} (\bibinfo {year} {2014})}\BibitemShut
  {NoStop}%
\bibitem [{\citenamefont {Berger}\ \emph {et~al.}(2011)\citenamefont {Berger},
  \citenamefont {Fennie},\ and\ \citenamefont {Neaton}}]{Berger2011}%
  \BibitemOpen
  \bibfield  {author} {\bibinfo {author} {\bibfnamefont {R.~F.}\ \bibnamefont
  {Berger}}, \bibinfo {author} {\bibfnamefont {C.~J.}\ \bibnamefont {Fennie}},
  \ and\ \bibinfo {author} {\bibfnamefont {J.~B.}\ \bibnamefont {Neaton}},\
  }\href {\doibase 10.1103/PhysRevLett.107.146804} {\bibfield  {journal}
  {\bibinfo  {journal} {Phys. Rev. Lett.}\ }\textbf {\bibinfo {volume} {107}},\
  \bibinfo {pages} {146804} (\bibinfo {year} {2011})}\BibitemShut {NoStop}%
\bibitem [{\citenamefont {Grote}\ and\ \citenamefont
  {Berger}(2015)}]{Grote2015}%
  \BibitemOpen
  \bibfield  {author} {\bibinfo {author} {\bibfnamefont {C.}~\bibnamefont
  {Grote}}\ and\ \bibinfo {author} {\bibfnamefont {R.~F.}\ \bibnamefont
  {Berger}},\ }\href {\doibase 10.1021/acs.jpcc.5b07446} {\bibfield  {journal}
  {\bibinfo  {journal} {J. Phys. Chem. C}\ }\textbf {\bibinfo {volume} {119}},\
  \bibinfo {pages} {22832} (\bibinfo {year} {2015})}\BibitemShut {NoStop}%
\bibitem [{\citenamefont {Teply}\ \emph {et~al.}(2021)\citenamefont {Teply},
  \citenamefont {Tyler},\ and\ \citenamefont {Berger}}]{Teply2021}%
  \BibitemOpen
  \bibfield  {author} {\bibinfo {author} {\bibfnamefont {C.}~\bibnamefont
  {Teply}}, \bibinfo {author} {\bibfnamefont {B.~A.}\ \bibnamefont {Tyler}}, \
  and\ \bibinfo {author} {\bibfnamefont {R.~F.}\ \bibnamefont {Berger}},\
  }\href {\doibase 10.1021/acs.jpcc.1c07169} {\bibfield  {journal} {\bibinfo
  {journal} {J. Phys. Chem. C}\ }\textbf {\bibinfo {volume} {125}},\ \bibinfo
  {pages} {25951} (\bibinfo {year} {2021})}\BibitemShut {NoStop}%
\bibitem [{\citenamefont {Callori}\ \emph {et~al.}(2015)\citenamefont
  {Callori}, \citenamefont {Hu}, \citenamefont {Bertinshaw}, \citenamefont
  {Yue}, \citenamefont {Danilkin}, \citenamefont {Wang}, \citenamefont
  {Nagarajan}, \citenamefont {Klose}, \citenamefont {Seidel},\ and\
  \citenamefont {Ulrich}}]{Callori2015}%
  \BibitemOpen
  \bibfield  {author} {\bibinfo {author} {\bibfnamefont {S.~J.}\ \bibnamefont
  {Callori}}, \bibinfo {author} {\bibfnamefont {S.}~\bibnamefont {Hu}},
  \bibinfo {author} {\bibfnamefont {J.}~\bibnamefont {Bertinshaw}}, \bibinfo
  {author} {\bibfnamefont {Z.~J.}\ \bibnamefont {Yue}}, \bibinfo {author}
  {\bibfnamefont {S.}~\bibnamefont {Danilkin}}, \bibinfo {author}
  {\bibfnamefont {X.~L.}\ \bibnamefont {Wang}}, \bibinfo {author}
  {\bibfnamefont {V.}~\bibnamefont {Nagarajan}}, \bibinfo {author}
  {\bibfnamefont {F.}~\bibnamefont {Klose}}, \bibinfo {author} {\bibfnamefont
  {J.}~\bibnamefont {Seidel}}, \ and\ \bibinfo {author} {\bibfnamefont
  {C.}~\bibnamefont {Ulrich}},\ }\href@noop {} {\bibfield  {journal} {\bibinfo
  {journal} {Phys. Rev. B}\ }\textbf {\bibinfo {volume} {91}},\ \bibinfo
  {pages} {140405} (\bibinfo {year} {2015})}\BibitemShut {NoStop}%
\bibitem [{\citenamefont {Lee}\ and\ \citenamefont {Rabe}(2011)}]{Lee2011}%
  \BibitemOpen
  \bibfield  {author} {\bibinfo {author} {\bibfnamefont {J.~H.}\ \bibnamefont
  {Lee}}\ and\ \bibinfo {author} {\bibfnamefont {K.~M.}\ \bibnamefont {Rabe}},\
  }\href@noop {} {\bibfield  {journal} {\bibinfo  {journal} {Phys. Rev. Lett.}\
  }\textbf {\bibinfo {volume} {107}},\ \bibinfo {pages} {067601} (\bibinfo
  {year} {2011})}\BibitemShut {NoStop}%
\bibitem [{\citenamefont {Paul}\ and\ \citenamefont
  {Birol}(2019)}]{Paul2019Vanadate}%
  \BibitemOpen
  \bibfield  {author} {\bibinfo {author} {\bibfnamefont {A.}~\bibnamefont
  {Paul}}\ and\ \bibinfo {author} {\bibfnamefont {T.}~\bibnamefont {Birol}},\
  }\href {\doibase 10.1103/PhysRevMaterials.3.085001} {\bibfield  {journal}
  {\bibinfo  {journal} {Phys. Rev. Mater.}\ }\textbf {\bibinfo {volume} {3}},\
  \bibinfo {pages} {085001} (\bibinfo {year} {2019})}\BibitemShut {NoStop}%
\bibitem [{\citenamefont {Kresse}\ and\ \citenamefont
  {Hafner}(1993)}]{Kresse1993}%
  \BibitemOpen
  \bibfield  {author} {\bibinfo {author} {\bibfnamefont {G.}~\bibnamefont
  {Kresse}}\ and\ \bibinfo {author} {\bibfnamefont {J.}~\bibnamefont
  {Hafner}},\ }\href@noop {} {\bibfield  {journal} {\bibinfo  {journal} {Phys.
  Rev. B}\ }\textbf {\bibinfo {volume} {47}},\ \bibinfo {pages} {558} (\bibinfo
  {year} {1993})}\BibitemShut {NoStop}%
\bibitem [{\citenamefont {Kresse}\ and\ \citenamefont
  {Joubert}(1999)}]{Kresse1999}%
  \BibitemOpen
  \bibfield  {author} {\bibinfo {author} {\bibfnamefont {G.}~\bibnamefont
  {Kresse}}\ and\ \bibinfo {author} {\bibfnamefont {D.}~\bibnamefont
  {Joubert}},\ }\href@noop {} {\bibfield  {journal} {\bibinfo  {journal} {Phys.
  Rev. B}\ }\textbf {\bibinfo {volume} {59}},\ \bibinfo {pages} {1758}
  (\bibinfo {year} {1999})}\BibitemShut {NoStop}%
\bibitem [{\citenamefont {Bl\"ochl}(1994)}]{Blochl1994}%
  \BibitemOpen
  \bibfield  {author} {\bibinfo {author} {\bibfnamefont {P.~E.}\ \bibnamefont
  {Bl\"ochl}},\ }\href@noop {} {\bibfield  {journal} {\bibinfo  {journal}
  {Phys. Rev. B}\ }\textbf {\bibinfo {volume} {50}},\ \bibinfo {pages} {17953}
  (\bibinfo {year} {1994})}\BibitemShut {NoStop}%
\bibitem [{\citenamefont {Perdew}\ \emph {et~al.}(2008)\citenamefont {Perdew},
  \citenamefont {Ruzsinszky}, \citenamefont {Csonka}, \citenamefont {Vydrov},
  \citenamefont {Scuseria}, \citenamefont {Constantin}, \citenamefont {Zhou},\
  and\ \citenamefont {Burke}}]{Perdew2008}%
  \BibitemOpen
  \bibfield  {author} {\bibinfo {author} {\bibfnamefont {J.~P.}\ \bibnamefont
  {Perdew}}, \bibinfo {author} {\bibfnamefont {A.}~\bibnamefont {Ruzsinszky}},
  \bibinfo {author} {\bibfnamefont {G.~I.}\ \bibnamefont {Csonka}}, \bibinfo
  {author} {\bibfnamefont {O.~A.}\ \bibnamefont {Vydrov}}, \bibinfo {author}
  {\bibfnamefont {G.~E.}\ \bibnamefont {Scuseria}}, \bibinfo {author}
  {\bibfnamefont {L.~A.}\ \bibnamefont {Constantin}}, \bibinfo {author}
  {\bibfnamefont {X.}~\bibnamefont {Zhou}}, \ and\ \bibinfo {author}
  {\bibfnamefont {K.}~\bibnamefont {Burke}},\ }\href@noop {} {\bibfield
  {journal} {\bibinfo  {journal} {Phys. Rev. Lett.}\ }\textbf {\bibinfo
  {volume} {100}},\ \bibinfo {pages} {136406} (\bibinfo {year}
  {2008})}\BibitemShut {NoStop}%
\bibitem [{\citenamefont {Togo}\ and\ \citenamefont {Tanaka}(2015)}]{Togo2015}%
  \BibitemOpen
  \bibfield  {author} {\bibinfo {author} {\bibfnamefont {A.}~\bibnamefont
  {Togo}}\ and\ \bibinfo {author} {\bibfnamefont {I.}~\bibnamefont {Tanaka}},\
  }\href@noop {} {\bibfield  {journal} {\bibinfo  {journal} {Scr. Mater.}\
  }\textbf {\bibinfo {volume} {108}},\ \bibinfo {pages} {1} (\bibinfo {year}
  {2015})}\BibitemShut {NoStop}%
\bibitem [{\citenamefont {Baroni}\ \emph {et~al.}(2001)\citenamefont {Baroni},
  \citenamefont {de~Gironcoli}, \citenamefont {Dal~Corso},\ and\ \citenamefont
  {Giannozzi}}]{Baroni2001}%
  \BibitemOpen
  \bibfield  {author} {\bibinfo {author} {\bibfnamefont {S.}~\bibnamefont
  {Baroni}}, \bibinfo {author} {\bibfnamefont {S.}~\bibnamefont
  {de~Gironcoli}}, \bibinfo {author} {\bibfnamefont {A.}~\bibnamefont
  {Dal~Corso}}, \ and\ \bibinfo {author} {\bibfnamefont {P.}~\bibnamefont
  {Giannozzi}},\ }\href@noop {} {\bibfield  {journal} {\bibinfo  {journal}
  {Rev. Mod. Phys.}\ }\textbf {\bibinfo {volume} {73}},\ \bibinfo {pages} {515}
  (\bibinfo {year} {2001})}\BibitemShut {NoStop}%
\bibitem [{\citenamefont {Campbell}\ \emph {et~al.}(2006)\citenamefont
  {Campbell}, \citenamefont {Stokes}, \citenamefont {Tanner},\ and\
  \citenamefont {Hatch}}]{Campbell2006}%
  \BibitemOpen
  \bibfield  {author} {\bibinfo {author} {\bibfnamefont {B.~J.}\ \bibnamefont
  {Campbell}}, \bibinfo {author} {\bibfnamefont {H.~T.}\ \bibnamefont
  {Stokes}}, \bibinfo {author} {\bibfnamefont {D.~E.}\ \bibnamefont {Tanner}},
  \ and\ \bibinfo {author} {\bibfnamefont {D.~M.}\ \bibnamefont {Hatch}},\
  }\href@noop {} {\bibfield  {journal} {\bibinfo  {journal} {J. Appl.
  Crystallogr.}\ }\textbf {\bibinfo {volume} {39}},\ \bibinfo {pages} {607}
  (\bibinfo {year} {2006})}\BibitemShut {NoStop}%
\bibitem [{\citenamefont {Aroyo}\ \emph {et~al.}(2014)\citenamefont {Aroyo},
  \citenamefont {Orobengoa}, \citenamefont {de~la Flor}, \citenamefont {Tasci},
  \citenamefont {Perez-Mato},\ and\ \citenamefont {Wondratschek}}]{Aroyo2014}%
  \BibitemOpen
  \bibfield  {author} {\bibinfo {author} {\bibfnamefont {M.~I.}\ \bibnamefont
  {Aroyo}}, \bibinfo {author} {\bibfnamefont {D.}~\bibnamefont {Orobengoa}},
  \bibinfo {author} {\bibfnamefont {G.}~\bibnamefont {de~la Flor}}, \bibinfo
  {author} {\bibfnamefont {E.~S.}\ \bibnamefont {Tasci}}, \bibinfo {author}
  {\bibfnamefont {J.~M.}\ \bibnamefont {Perez-Mato}}, \ and\ \bibinfo {author}
  {\bibfnamefont {H.}~\bibnamefont {Wondratschek}},\ }\href@noop {} {\bibfield
  {journal} {\bibinfo  {journal} {Acta Crystallogr. A}\ }\textbf {\bibinfo
  {volume} {70}},\ \bibinfo {pages} {126} (\bibinfo {year} {2014})}\BibitemShut
  {NoStop}%
\bibitem [{\citenamefont {Momma}\ and\ \citenamefont
  {Izumi}(2011)}]{Momma2011}%
  \BibitemOpen
  \bibfield  {author} {\bibinfo {author} {\bibfnamefont {K.}~\bibnamefont
  {Momma}}\ and\ \bibinfo {author} {\bibfnamefont {F.}~\bibnamefont {Izumi}},\
  }\href@noop {} {\bibfield  {journal} {\bibinfo  {journal} {J. Appl.
  Crystallogr.}\ }\textbf {\bibinfo {volume} {44}},\ \bibinfo {pages} {1272}
  (\bibinfo {year} {2011})}\BibitemShut {NoStop}%
\bibitem [{\citenamefont {Yun}\ \emph {et~al.}(2022)\citenamefont {Yun},
  \citenamefont {Gautreau}, \citenamefont {Mkhoyan},\ and\ \citenamefont
  {Birol}}]{Yun2022}%
  \BibitemOpen
  \bibfield  {author} {\bibinfo {author} {\bibfnamefont {H.}~\bibnamefont
  {Yun}}, \bibinfo {author} {\bibfnamefont {D.}~\bibnamefont {Gautreau}},
  \bibinfo {author} {\bibfnamefont {K.~A.}\ \bibnamefont {Mkhoyan}}, \ and\
  \bibinfo {author} {\bibfnamefont {T.}~\bibnamefont {Birol}},\ }\href@noop {}
  {\enquote {\bibinfo {title} {https://hdl.handle.net/11299/241504},}\ }
  (\bibinfo {year} {2022})\BibitemShut {NoStop}%
\bibitem [{\citenamefont {De~Graef}\ and\ \citenamefont
  {McHenry}(2012)}]{Graef2012}%
  \BibitemOpen
  \bibfield  {author} {\bibinfo {author} {\bibfnamefont {M.}~\bibnamefont
  {De~Graef}}\ and\ \bibinfo {author} {\bibfnamefont {M.~E.}\ \bibnamefont
  {McHenry}},\ }\href {\doibase 10.1017/CBO9781139051637} {\emph {\bibinfo
  {title} {Structure of Materials: An Introduction to Crystallography,
  Diffraction and Symmetry}}},\ \bibinfo {edition} {2nd}\ ed.\ (\bibinfo
  {publisher} {Cambridge University Press},\ \bibinfo {year}
  {2012})\BibitemShut {NoStop}%
\bibitem [{\citenamefont {Hatch}\ and\ \citenamefont
  {Stokes}(1987)}]{Hatch1987}%
  \BibitemOpen
  \bibfield  {author} {\bibinfo {author} {\bibfnamefont {D.~M.}\ \bibnamefont
  {Hatch}}\ and\ \bibinfo {author} {\bibfnamefont {H.~T.}\ \bibnamefont
  {Stokes}},\ }\href {\doibase 10.1103/PhysRevB.35.8509} {\bibfield  {journal}
  {\bibinfo  {journal} {Phys. Rev. B}\ }\textbf {\bibinfo {volume} {35}},\
  \bibinfo {pages} {8509} (\bibinfo {year} {1987})}\BibitemShut {NoStop}%
\bibitem [{\citenamefont {Hatch}\ \emph {et~al.}(1989)\citenamefont {Hatch},
  \citenamefont {Stokes}, \citenamefont {Aleksandrov},\ and\ \citenamefont
  {Misyul}}]{Hatch1989}%
  \BibitemOpen
  \bibfield  {author} {\bibinfo {author} {\bibfnamefont {D.~M.}\ \bibnamefont
  {Hatch}}, \bibinfo {author} {\bibfnamefont {H.~T.}\ \bibnamefont {Stokes}},
  \bibinfo {author} {\bibfnamefont {K.~S.}\ \bibnamefont {Aleksandrov}}, \ and\
  \bibinfo {author} {\bibfnamefont {S.~V.}\ \bibnamefont {Misyul}},\ }\href
  {\doibase 10.1103/PhysRevB.39.9282} {\bibfield  {journal} {\bibinfo
  {journal} {Phys. Rev. B}\ }\textbf {\bibinfo {volume} {39}},\ \bibinfo
  {pages} {9282} (\bibinfo {year} {1989})}\BibitemShut {NoStop}%
\bibitem [{SM()}]{SM}%
  \BibitemOpen
  \bibfield {title} {\bibinfo {title} {See Supplemental Material at [URL]
  for atomic model of the experimental structure, brillouin zone of a
  body-centered tetragonal structure, atomic displacement associated with the
  imaginary frequencies at the P- and N- points, atomic configurations of the x
  modes, order parameter directions of the X irreps, bond valence and
  visualization of cation polyhedral, GII calculations, strain effect on the
  lattice dynamics, Landau analysis of the X modes of the I4/mmm Sr$_2$SnO$_4$
  structure, strain response of OPD configurations of the X modes, strain
  effect on the ground-state structure, electronic structure, pressure
  response of Sr$_2$SnO$_4$}}\ \BibitemShut {NoStop}%
\bibitem [{\citenamefont {Hatch}\ and\ \citenamefont
  {Stokes}(2003)}]{Hatch2003}%
  \BibitemOpen
  \bibfield  {author} {\bibinfo {author} {\bibfnamefont {D.~M.}\ \bibnamefont
  {Hatch}}\ and\ \bibinfo {author} {\bibfnamefont {H.~T.}\ \bibnamefont
  {Stokes}},\ }\href {\doibase 10.1107/s0021889803005946} {\bibfield  {journal}
  {\bibinfo  {journal} {J. Appl. Crystallogr.}\ }\textbf {\bibinfo {volume}
  {36}},\ \bibinfo {pages} {951} (\bibinfo {year} {2003})}\BibitemShut
  {NoStop}%
\bibitem [{\citenamefont {Brown}(2009)}]{Brown2009}%
  \BibitemOpen
  \bibfield  {author} {\bibinfo {author} {\bibfnamefont {I.~D.}\ \bibnamefont
  {Brown}},\ }\href@noop {} {\bibfield  {journal} {\bibinfo  {journal} {Chem.
  Rev.}\ }\textbf {\bibinfo {volume} {109}},\ \bibinfo {pages} {6858} (\bibinfo
  {year} {2009})}\BibitemShut {NoStop}%
\bibitem [{\citenamefont {Brown}(1978)}]{Brown1978}%
  \BibitemOpen
  \bibfield  {author} {\bibinfo {author} {\bibfnamefont {I.~D.}\ \bibnamefont
  {Brown}},\ }\href@noop {} {\bibfield  {journal} {\bibinfo  {journal} {Chem.
  Soc. Rev.}\ }\textbf {\bibinfo {volume} {7}},\ \bibinfo {pages} {359}
  (\bibinfo {year} {1978})}\BibitemShut {NoStop}%
\bibitem [{\citenamefont {Salinas-Sanchez}\ \emph {et~al.}(1992)\citenamefont
  {Salinas-Sanchez}, \citenamefont {Garcia-Muñoz}, \citenamefont
  {Rodriguez-Carvajal}, \citenamefont {Saez-Puche},\ and\ \citenamefont
  {Martinez}}]{Salinas-Sanchez1992}%
  \BibitemOpen
  \bibfield  {author} {\bibinfo {author} {\bibfnamefont {A.}~\bibnamefont
  {Salinas-Sanchez}}, \bibinfo {author} {\bibfnamefont {J.}~\bibnamefont
  {Garcia-Muñoz}}, \bibinfo {author} {\bibfnamefont {J.}~\bibnamefont
  {Rodriguez-Carvajal}}, \bibinfo {author} {\bibfnamefont {R.}~\bibnamefont
  {Saez-Puche}}, \ and\ \bibinfo {author} {\bibfnamefont {J.}~\bibnamefont
  {Martinez}},\ }\href@noop {} {\bibfield  {journal} {\bibinfo  {journal} {J.
  Solid State Chem.}\ }\textbf {\bibinfo {volume} {100}},\ \bibinfo {pages}
  {201} (\bibinfo {year} {1992})}\BibitemShut {NoStop}%
\bibitem [{\citenamefont {Brese}\ and\ \citenamefont
  {O'Keeffe}(1991)}]{Brese1991}%
  \BibitemOpen
  \bibfield  {author} {\bibinfo {author} {\bibfnamefont {N.~E.}\ \bibnamefont
  {Brese}}\ and\ \bibinfo {author} {\bibfnamefont {M.}~\bibnamefont
  {O'Keeffe}},\ }\href {\doibase 10.1107/S0108768190011041} {\bibfield
  {journal} {\bibinfo  {journal} {Acta Crystallogr. B}\ }\textbf {\bibinfo
  {volume} {47}},\ \bibinfo {pages} {192} (\bibinfo {year} {1991})}\BibitemShut
  {NoStop}%
\bibitem [{\citenamefont {Woodward}(1997{\natexlab{b}})}]{Woodward1997b}%
  \BibitemOpen
  \bibfield  {author} {\bibinfo {author} {\bibfnamefont {P.~M.}\ \bibnamefont
  {Woodward}},\ }\href {\doibase 10.1107/S0108768196012050} {\bibfield
  {journal} {\bibinfo  {journal} {Acta Crystallogr. B}\ }\textbf {\bibinfo
  {volume} {53}},\ \bibinfo {pages} {44} (\bibinfo {year}
  {1997}{\natexlab{b}})}\BibitemShut {NoStop}%
\bibitem [{\citenamefont {Haeni}\ \emph {et~al.}(2001)\citenamefont {Haeni},
  \citenamefont {Theis}, \citenamefont {Schlom}, \citenamefont {Tian},
  \citenamefont {Pan}, \citenamefont {Chang}, \citenamefont {Takeuchi},\ and\
  \citenamefont {Xiang}}]{Haeni2001}%
  \BibitemOpen
  \bibfield  {author} {\bibinfo {author} {\bibfnamefont {J.~H.}\ \bibnamefont
  {Haeni}}, \bibinfo {author} {\bibfnamefont {C.~D.}\ \bibnamefont {Theis}},
  \bibinfo {author} {\bibfnamefont {D.~G.}\ \bibnamefont {Schlom}}, \bibinfo
  {author} {\bibfnamefont {W.}~\bibnamefont {Tian}}, \bibinfo {author}
  {\bibfnamefont {X.~Q.}\ \bibnamefont {Pan}}, \bibinfo {author} {\bibfnamefont
  {H.}~\bibnamefont {Chang}}, \bibinfo {author} {\bibfnamefont
  {I.}~\bibnamefont {Takeuchi}}, \ and\ \bibinfo {author} {\bibfnamefont
  {X.-D.}\ \bibnamefont {Xiang}},\ }\href {\doibase 10.1063/1.1371788}
  {\bibfield  {journal} {\bibinfo  {journal} {Appl. Phys. Lett.}\ }\textbf
  {\bibinfo {volume} {78}},\ \bibinfo {pages} {3292} (\bibinfo {year}
  {2001})}\BibitemShut {NoStop}%
\bibitem [{\citenamefont {Burganov}\ \emph {et~al.}(2016)\citenamefont
  {Burganov}, \citenamefont {Adamo}, \citenamefont {Mulder}, \citenamefont
  {Uchida}, \citenamefont {King}, \citenamefont {Harter}, \citenamefont {Shai},
  \citenamefont {Gibbs}, \citenamefont {Mackenzie}, \citenamefont {Uecker},
  \citenamefont {Bruetzam}, \citenamefont {Beasley}, \citenamefont {Fennie},
  \citenamefont {Schlom},\ and\ \citenamefont {Shen}}]{Burganov2016}%
  \BibitemOpen
  \bibfield  {author} {\bibinfo {author} {\bibfnamefont {B.}~\bibnamefont
  {Burganov}}, \bibinfo {author} {\bibfnamefont {C.}~\bibnamefont {Adamo}},
  \bibinfo {author} {\bibfnamefont {A.}~\bibnamefont {Mulder}}, \bibinfo
  {author} {\bibfnamefont {M.}~\bibnamefont {Uchida}}, \bibinfo {author}
  {\bibfnamefont {P.~D.~C.}\ \bibnamefont {King}}, \bibinfo {author}
  {\bibfnamefont {J.~W.}\ \bibnamefont {Harter}}, \bibinfo {author}
  {\bibfnamefont {D.~E.}\ \bibnamefont {Shai}}, \bibinfo {author}
  {\bibfnamefont {A.~S.}\ \bibnamefont {Gibbs}}, \bibinfo {author}
  {\bibfnamefont {A.~P.}\ \bibnamefont {Mackenzie}}, \bibinfo {author}
  {\bibfnamefont {R.}~\bibnamefont {Uecker}}, \bibinfo {author} {\bibfnamefont
  {M.}~\bibnamefont {Bruetzam}}, \bibinfo {author} {\bibfnamefont {M.~R.}\
  \bibnamefont {Beasley}}, \bibinfo {author} {\bibfnamefont {C.~J.}\
  \bibnamefont {Fennie}}, \bibinfo {author} {\bibfnamefont {D.~G.}\
  \bibnamefont {Schlom}}, \ and\ \bibinfo {author} {\bibfnamefont {K.~M.}\
  \bibnamefont {Shen}},\ }\href {\doibase 10.1103/PhysRevLett.116.197003}
  {\bibfield  {journal} {\bibinfo  {journal} {Phys. Rev. Lett.}\ }\textbf
  {\bibinfo {volume} {116}},\ \bibinfo {pages} {197003} (\bibinfo {year}
  {2016})}\BibitemShut {NoStop}%
\bibitem [{\citenamefont {Singh}\ \emph {et~al.}(2014)\citenamefont {Singh},
  \citenamefont {Xu},\ and\ \citenamefont {Ong}}]{Singh2014}%
  \BibitemOpen
  \bibfield  {author} {\bibinfo {author} {\bibfnamefont {D.~J.}\ \bibnamefont
  {Singh}}, \bibinfo {author} {\bibfnamefont {Q.}~\bibnamefont {Xu}}, \ and\
  \bibinfo {author} {\bibfnamefont {K.~P.}\ \bibnamefont {Ong}},\ }\href@noop
  {} {\bibfield  {journal} {\bibinfo  {journal} {Appl. Phys. Lett.}\ }\textbf
  {\bibinfo {volume} {104}},\ \bibinfo {pages} {011910} (\bibinfo {year}
  {2014})}\BibitemShut {NoStop}%
\bibitem [{\citenamefont {Gao}\ \emph {et~al.}(2020)\citenamefont {Gao},
  \citenamefont {Li}, \citenamefont {Zhao}, \citenamefont {Lv}, \citenamefont
  {Li}, \citenamefont {Zhang}, \citenamefont {Du},\ and\ \citenamefont
  {Liu}}]{Gao2020}%
  \BibitemOpen
  \bibfield  {author} {\bibinfo {author} {\bibfnamefont {Q.}~\bibnamefont
  {Gao}}, \bibinfo {author} {\bibfnamefont {K.}~\bibnamefont {Li}}, \bibinfo
  {author} {\bibfnamefont {L.}~\bibnamefont {Zhao}}, \bibinfo {author}
  {\bibfnamefont {K.}~\bibnamefont {Lv}}, \bibinfo {author} {\bibfnamefont
  {H.}~\bibnamefont {Li}}, \bibinfo {author} {\bibfnamefont {J.}~\bibnamefont
  {Zhang}}, \bibinfo {author} {\bibfnamefont {W.}~\bibnamefont {Du}}, \ and\
  \bibinfo {author} {\bibfnamefont {Q.}~\bibnamefont {Liu}},\ }\href@noop {}
  {\bibfield  {journal} {\bibinfo  {journal} {J. Mater. Chem. C}\ }\textbf
  {\bibinfo {volume} {8}},\ \bibinfo {pages} {3545} (\bibinfo {year}
  {2020})}\BibitemShut {NoStop}%
\bibitem [{\citenamefont {Baniecki}\ \emph {et~al.}(2017)\citenamefont
  {Baniecki}, \citenamefont {Yamazaki}, \citenamefont {Ricinschi},
  \citenamefont {{Van Overmeere}}, \citenamefont {Aso}, \citenamefont {Miyata},
  \citenamefont {Yamada}, \citenamefont {Fujimura}, \citenamefont {Maran},
  \citenamefont {Anazawa}, \citenamefont {Valanoor},\ and\ \citenamefont
  {Imanaka}}]{Baniecki2017}%
  \BibitemOpen
  \bibfield  {author} {\bibinfo {author} {\bibfnamefont {J.~D.}\ \bibnamefont
  {Baniecki}}, \bibinfo {author} {\bibfnamefont {T.}~\bibnamefont {Yamazaki}},
  \bibinfo {author} {\bibfnamefont {D.}~\bibnamefont {Ricinschi}}, \bibinfo
  {author} {\bibfnamefont {Q.}~\bibnamefont {{Van Overmeere}}}, \bibinfo
  {author} {\bibfnamefont {H.}~\bibnamefont {Aso}}, \bibinfo {author}
  {\bibfnamefont {Y.}~\bibnamefont {Miyata}}, \bibinfo {author} {\bibfnamefont
  {H.}~\bibnamefont {Yamada}}, \bibinfo {author} {\bibfnamefont
  {N.}~\bibnamefont {Fujimura}}, \bibinfo {author} {\bibfnamefont
  {R.}~\bibnamefont {Maran}}, \bibinfo {author} {\bibfnamefont
  {T.}~\bibnamefont {Anazawa}}, \bibinfo {author} {\bibfnamefont
  {N.}~\bibnamefont {Valanoor}}, \ and\ \bibinfo {author} {\bibfnamefont
  {Y.}~\bibnamefont {Imanaka}},\ }\href {\doibase 10.1038/srep41725} {\bibfield
   {journal} {\bibinfo  {journal} {Sci. Rep.}\ }\textbf {\bibinfo {volume}
  {7}},\ \bibinfo {pages} {41725} (\bibinfo {year} {2017})}\BibitemShut
  {NoStop}%
\bibitem [{\citenamefont {Prakash}\ and\ \citenamefont
  {Jalan}(2019)}]{Prakash2019}%
  \BibitemOpen
  \bibfield  {author} {\bibinfo {author} {\bibfnamefont {A.}~\bibnamefont
  {Prakash}}\ and\ \bibinfo {author} {\bibfnamefont {B.}~\bibnamefont
  {Jalan}},\ }\href {\doibase 10.1002/admi.201900479} {\bibfield  {journal}
  {\bibinfo  {journal} {Adv. Mater. Interfaces}\ }\textbf {\bibinfo {volume}
  {6}},\ \bibinfo {pages} {1} (\bibinfo {year} {2019})}\BibitemShut {NoStop}%
\bibitem [{\citenamefont {He}\ \emph {et~al.}(2020)\citenamefont {He},
  \citenamefont {Yang}, \citenamefont {Xu}, \citenamefont {Smith},
  \citenamefont {Yang},\ and\ \citenamefont {Sun}}]{He2020}%
  \BibitemOpen
  \bibfield  {author} {\bibinfo {author} {\bibfnamefont {H.}~\bibnamefont
  {He}}, \bibinfo {author} {\bibfnamefont {Z.}~\bibnamefont {Yang}}, \bibinfo
  {author} {\bibfnamefont {Y.}~\bibnamefont {Xu}}, \bibinfo {author}
  {\bibfnamefont {A.~T.}\ \bibnamefont {Smith}}, \bibinfo {author}
  {\bibfnamefont {G.}~\bibnamefont {Yang}}, \ and\ \bibinfo {author}
  {\bibfnamefont {L.}~\bibnamefont {Sun}},\ }\href {\doibase
  10.1186/s40580-020-00242-7} {\bibfield  {journal} {\bibinfo  {journal} {Nano
  Converg.}\ }\textbf {\bibinfo {volume} {7}},\ \bibinfo {pages} {32} (\bibinfo
  {year} {2020})}\BibitemShut {NoStop}%
\bibitem [{\citenamefont {Ong}\ \emph {et~al.}(2015)\citenamefont {Ong},
  \citenamefont {Fan}, \citenamefont {Subedi}, \citenamefont {Sullivan},\ and\
  \citenamefont {Singh}}]{Khuong2015}%
  \BibitemOpen
  \bibfield  {author} {\bibinfo {author} {\bibfnamefont {K.~P.}\ \bibnamefont
  {Ong}}, \bibinfo {author} {\bibfnamefont {X.}~\bibnamefont {Fan}}, \bibinfo
  {author} {\bibfnamefont {A.}~\bibnamefont {Subedi}}, \bibinfo {author}
  {\bibfnamefont {M.~B.}\ \bibnamefont {Sullivan}}, \ and\ \bibinfo {author}
  {\bibfnamefont {D.~J.}\ \bibnamefont {Singh}},\ }\href {\doibase
  10.1063/1.4919564} {\bibfield  {journal} {\bibinfo  {journal} {APL Mater.}\
  }\textbf {\bibinfo {volume} {3}},\ \bibinfo {pages} {062505} (\bibinfo {year}
  {2015})}\BibitemShut {NoStop}%
\end{thebibliography}
\end{document}